\documentclass[a4paper,12pt]{article}
\usepackage{amsmath,amsfonts,amssymb,cite}
\usepackage{graphicx}
\makeatletter
\newcommand{\l@abcd}[2]{\hbox{} \hbox to\textwidth{#1\hfil #2}}
\makeatother

\numberwithin{equation}{section}

\usepackage{hyperref}

\usepackage[dvipsnames]{xcolor}

\newcommand{\lb}[0]{\left(}
\newcommand{\rb}[0]{\right)}
\newcommand{\lsb}[0]{\left[}
\newcommand{\rsb}[0]{\right]}
\newcommand{\eqzz}[0]{\underset{z\to0}{=}}

\newcommand{\vk}[0]{\varkappa}
\newcommand{\pz}[0]{\partial_z}

\allowdisplaybreaks

\begin{document}

\begin{center}
{\large\bf Towards a theory of bottom-up holographic models for linear Regge trajectories of light mesons}
\end{center}
\bigskip
\begin{center}
{ S. S. Afonin
and T. D. Solomko}
\end{center}

\begin{center}
{\small\it Saint Petersburg State University, 7/9 Universitetskaya nab.,
St.Petersburg, 199034, Russia}
\end{center}

\bigskip

\begin{abstract}
We advance in constructing a bottom-up holographic theory of linear meson Regge trajectories that generalizes and unites into one logical framework various
bottom-up holographic approaches proposed in the past and scattered in the literature.
The starting point of the theory is a quadratic in fields holographic five-dimensional action in which the Poincar\'{e} invariance
along the holographic coordinate is violated in the most general way compatible with the linear Regge behavior of the discrete spectrum in four dimensions.
It is further demonstrated how different Soft Wall (SW) like holographic models existing in the literature plus some new ones emerge from our general setup.
Various interrelations between the emerging models are studied. These models include the known SW models with different sign in the exponential background,
the SW models with certain generalized backgrounds, with modified metrics, and No Wall models with 5D mass depending on the holographic coordinate
in a simple polynomial way. We argue that this dependence allows to describe the effects caused by the main non-local phenomena of strongly coupled 4D gauge theory,
the confinement and chiral symmetry breaking, in terms of a local 5D dual field theory in the AdS space.
We provide a detailed comparison of our approach with the Light Front holographic QCD,
with the spectroscopic predictions of the dual Veneziano like amplitudes, and with the experimental Regge phenomenology.
We apply our general approach to a holographic study of confinement, chiral symmetry breaking, and the pion form factor.
\end{abstract}

\newpage

\tableofcontents
\newpage

\section{Introduction}

The modern bottom-up holographic QCD (AdS/QCD) represents a large set of phenomenological approaches inspired by the gauge/gravity duality in string theory~\cite{mald,witten,gub} and
applies the holographic methods developed for conformal field theories to the case of real QCD.
The holographic approach is based on a conjecture that observables in strongly coupled gauge theories, in the limit of a large number of colors, can be determined from classical  fields  weakly coupled through gravity in an Anti-de Sitter (AdS) space having one extra  dimension. The holographic approach to strong interactions includes also various top-down
holographic models which start from some brane construction within a string theory and try to get a dual model useful for the QCD phenomenology (see, e.g.,~\cite{kruc,bab,sakai,kruc2}). A natural implementation of Regge behavior in the hadron spectra and correct Operator Product Expansion (OPE) of correlation functions in QCD, which is in the focus of the present work, have still not been achieved in the top-down approach, for this reason the top-down holographic models will not be discussed.

The first elements of AdS/QCD approach appeared in the attempts to describe the glueball scattering and spectrum using the methods of AdS/CFT correspondence~\cite{pol1,pol2,BF1,BF2}.
Soon after that the approach was finally formulated in~\cite{son1,pom,Hirn:2005nr} and applied to description of spontaneous chiral symmetry breaking and spectrum of light mesons.
The incorporation of Chern-Simons term allowed to describe baryons and the physics related to the QCD chiral anomaly~\cite{Pomarol:2008aa}.
The overall agreement of the bottom-up holographic approach with the existing hadron phenomenology turned out to be surprisingly good.
This marked the birth of a new class of models describing the low-energy QCD and hadron spectra
with an accuracy comparable with old traditional approaches (effective field theories, potential quark models, etc.). Since that time (2005)
a great number of various bottom-up holographic models for strongly coupled QCD have been proposed and applied to description of the hadron phenomenology.
The number of papers on this subject is enormous,
among the most recent developments one can mention the construction of various new models partly describing
the hadron spectrum~\cite{AS,vento2021,vento2020,vento2018,afonin2020,Capossoli2019,Contreras2018,nonlSW,Zou:2019tpo,Ferreira:2019inu,Amorim:2021gat},
hadron structure~\cite{zahed3,contreras2021,Mamo:2021krl,Lyubovitskij:2020xqj,Lyubovitskij:2020gjz,Lyubovitskij:2020otz,EKpion,deTeramond:2018ecg,Bacchetta:2017vzh}
including the deep inelastic scattering~\cite{zahed2,Amorim:2021gat,Amorim:2021ffr,Watanabe:2019zny,zahed,Mamo:2018eoy}, and a very large field of QCD thermodynamics~\cite{Caldeira:2020rir,Caldeira:2020sot,MartinContreras:2021bis,Gutsche2020,FolcoCapossoli:2020pks,Ballon-Bayona:2020xls,Afonin2020PLB,Katanaeva2020,Katanaeva2019,Gutsche:2019blp,Gutsche:2019pls,Lv:2018wfq} in which even some elements of information theory can be exploited (see, e.g., the recent Refs.~\cite{daRocha:2021ntm,Braga:2021zyi} and references therein).
The important recent developments also include holographic modeling of dense baryonic matter for neutron stars (see the review~\cite{Jarvinen:2021jbd}), holographic modeling of hadronic light-by-light scattering for the muon anomalous magnetic moment (reviewed in~\cite{Leutgeb:2021bpo}), and some applications of bottom-up holographic setup borrowed from the hadron physics to other fields, such as the composite Higgs models~\cite{EKHiggs} and the high-temperature superconductivity~\cite{AP2016}. An interesting and quite fruitful branch of the AdS/QCD approach is the Light-Front (LF) holographic QCD~\cite{br3}
in which the holographic correspondence between the fields of a dual 5-dimensional theory and those of the 4-dimensional theory is realized at fixed light-front time.

Most of the aforementioned AdS/QCD models were built on the base
of the so-called Soft Wall (SW) holographic model introduced in Refs.~\cite{son2,andreev}. The mass scale $c$ is incorporated into these models via the exponential
scale factor $e^{cz^2}$ either in a 5D action of the dual theory~\cite{son2} (and should be then regarded as a part of Lagrangian) or in the 5D metric of AdS$_5$
space~\cite{andreev}. Here $z$ is the fifth coordinate called holographic which is interpreted as the inverse energy scale
(related to the quark-antiquark separation in the LF holographic QCD~\cite{br3}). The given scale factor is often called ``dilaton background'' or just
``background'' but it must be emphasized that this is a slang (that will be used in the present paper as well). This ``background'' should have a dynamical origin,
in particular, it was suggested to be a result of a closed string tachyon condensation in the original paper~\cite{son2}.
We are not aware of any explicit realization of this proposal but a bottom-up holographic model based on an open string tachyon condensation
(adopted in a simplified form from the string theory) was constructed in Ref.~\cite{Casero:2007ae} and worked out further in
Refs.~\cite{Gursoy:2007cb,Gursoy:2007er,Iatrakis:2010jb}.
Remarkably, the setup introduced in~\cite{Casero:2007ae} describes both the chiral symmetry breaking and asymptotically linear radial Regge trajectories.


The original SW model was designed to
describe the phenomenology of linear Regge trajectories in the large-$N_c$ limit of QCD~\cite{hoof,wit} but it turned out successful in other areas of hadron phenomenology,
in many cases demonstrating an intelligible interpolation between the low and high energy sectors of QCD.
The simplest SW holographic model can be viewed as the most self-consistent way of rewriting the infinite number of pole terms
(expected in the large-$N_c$ limit of QCD~\cite{hoof,wit}) with linear spectrum of masses squared,
in the pole representation of two-point Correlation Functions (CFs), as some 5D gravitational model of free fields~\cite{afonin2010,ahep}.
Remarkably, the holographic recipe of Refs.~\cite{witten,gub} for calculation of CFs follows in a natural way within such a rewriting.
This means, in particular, that the SW holographic models are closely related
with the planar QCD sum rules (in a sense, they represent just 5D rewriting of those sum rules~\cite{afonin2010})
which were widely used in the past to study the phenomenology of linear radial trajectories in the meson sector~\cite{lin2c}.
In descriptions of hadron electromagnetic form factors, the holographic approach, especially the SW one,
recovers the old pre-QCD dual description with all its phenomenological successes~\cite{zahed3}.
But the holographic QCD is much wider in scope --- its strong advantage consists in the use of
Lagrangian formulation that enables more refined calculations and opens the door to many other applications.

A top-down derivation of SW like holographic models from some brane construction in a string theory remains an open problem.
This problem, however, is purely theoretical and does not impede in building a rich holographic phenomenology.
It should be recalled that a similar situation persists in non-perturbative QCD --- there are various popular phenomenological models
for low-energy strong interactions
but no one of them has been derived from QCD. We believe that the bottom-up holographic models will remain useful and actively explored even if
in future somebody proves rigorously that the gauge/gravity duality cannot work for non-conformal theories. Such a result would make outdated
the holographic top-down approach and likely some bottom-up models but the SW (and likely some elements of Hard Wall) approach would
survive because, as we mentioned above, it became a useful language unifying into one logical framework many elements of various old
phenomenological approaches (QCD sum rules in the large-$N_c$ limit, light-cone QCD, Regge physics, deep inelastic scattering, chiral perturbation theory)
and reproducing many results from those approaches. For this reason, the further development and accurate formulization of this 5D language
for hadron physics looks rewarding.

At present there is no systematics in the existing abundance of various SW-like models. The numerous proposed modifications of the SW holographic model are usually aimed at
improvement of phenomenological description in some specific problem and the question how a proposed modification will work in other places often remains unaddressed.
For instance, many modifications suggested to improve agreement with the experimental spectroscopy in some sector, in reality would lead to
inadmissible analytical properties of the corresponding correlation functions, if those functions were calculated. One should not forget that the primary outcome
of the holographic approach is given by the CFs which replace observables in the conformal field theories. The mass spectrum represents
a by-product. Using the known theorem on the spectral decomposition of the Green functions (the two-point CFs) one can bypass the calculation of a CF
by finding the discrete spectrum from the corresponding equation of motion, as is usually implemented in practice. But this should not depreciate the importance
of respecting the correct analytical properties of underlying CF. This aspect becomes especially problematic when non-linear corrections to a Regge like spectrum
are introduced, whether manually or via a back-reaction of fields in dynamical AdS/QCD models.

The numerous practical applications show usually that the approximation of a static dilaton background and of probe limit
(i.e., when the 5D metrics is not back-reacted neither by a dilaton background nor by the matter fields)
is more than enough for phenomenological purposes.
Furthermore, we are not aware of any bottom-up holographic model beyond this approximation or a top-down holographic model for QCD
that would reproduce correctly the analytical structure of OPE of correlation functions in QCD (i.e., the perturbative logarithm plus power corrections).
Nevertheless, the construction of dynamical dilaton-gravity AdS/QCD models looks attracting theoretically.
The study of such dynamical models started in~\cite{Casero:2007ae,Gursoy:2007cb,Gursoy:2007er,Csaki:2006ji}
and was followed by many papers along this direction. In particular, one can fine-tune the dilaton potential in such a way that
a SW-like metric is reproduced~\cite{Casero:2007ae,Gursoy:2007er}. The dynamical AdS/QCD models are more complicated than models with a static
background and may look more appealing theoretically. However, they are not as successful in the QCD phenomenology, neither in describing experimental
data in hadron physics nor in reproducing various old-known relations from other approaches to strong interactions. It should be stressed
that the bottom-up AdS/QCD is a phenomenology-driven approach, so the agreement with the known phenomenology should be in first place in any judgement
about ``correctness'' of a model.

From a conceptual viewpoint, the consistency of dynamical holographic models with back-reaction is questionable
if the whole approach somehow follows from an underlying string theory.
Indeed, both the gravitational metric and dilaton background are then determined by underlying string dynamics,
hence, the metric is back-reacted by dilaton (and vice versa) indirectly, via this string dynamics, i.e., the given back-reaction cannot be fully
described just by a set of coupled Einstein equations for the metric and dilaton. An instructive example of this point is given
by an extensive analysis of Refs.~\cite{Casero:2007ae,Gursoy:2007cb,Gursoy:2007er,Iatrakis:2010jb}, where it was advocated that in exploring improved holographic theories for QCD
a seminal direction is to think of the 5D bulk theory as a (non-critical) string theory, not just gravity. It turns out that
it is a gaussian potential of scalar tachyon field that can give rise to a SW like background.
We see thus that the assumption that the underlying string dynamics can be
neglected in such a way that some effective dynamical dilaton persists, as a matter of fact, looks almost as poorly substantiated theoretically
as the assumption of effective static dilaton background. As long as the underlying string dynamics is unknown,
both the dynamic and static dilaton background represent just working hypotheses for building bottom-up models. Only the phenomenology can discriminate
which hypothesis works better, hence, is more "right". To the best of our knowledge, the most successful holographic model describing the Regge physics,
OPE of CFs, and hadron form-factors is the static SW holographic model. There is no purely theoretical justification
for this observation. In the pioneering paper~\cite{son2}, it was assumed that it is a closed string tachyon
condensation that should lead to a static dilaton background. The form of this background is dictated by the QCD phenomenology.


In the present paper, we develop a general theory of bottom-up holographic models describing the linear Regge and radial trajectories.
The starting point of our theory will be the most general quadratic in fields holographic 5D action violating the Poincar\'{e} invariance
along the holographic coordinate $z$ but which is compatible with the exactly linear Regge behavior of the mass spectrum.
After that we demonstrate how various SW like holographic models existing in the literature plus some new ones emerge from our general setup
and study interrelations between the emerging models.
We make a detailed comparison of the standard SW holographic approach with the LF holographic QCD,
with the spectroscopic predictions of the dual Veneziano like amplitudes, and with the experimental Regge phenomenology.
The correct analytical properties of CFs are guaranteed, the only troubling point appears in the case of vector correlation function,
where a unphysical massless pole arises in the general case. We construct a renormalization recipe that allows to avoid this problem and even,
in a certain variant, to predict the intercept of linear radial trajectory. The absence of massless pole in the original SW model~\cite{son2}
turns out to be a particular case of this recipe which takes place for a particular formulation of the model. We apply our general approach
to a holographic study of confinement, chiral symmetry breaking, and the pion form factor. We demonstrate, in particular, that the exact vector
meson dominance for the pion form factor holds only in the case of the SW model with negative exponential background $e^{-|c|z^2}$, within other formulations,
all radially excited states contribute.

The bottom-up holographic theory that we develop in the present paper encompasses and generalizes various ideas proposed in the past
and scattered in the literature. Such an extensive analysis is performed for the first time since the appearance of SW models in the pioneering paper~\cite{son2}.
The original SW holographic model of Ref.~\cite{son2} with the exponential background $e^{-|c|z^2}$
was rapidly followed by numerous extensions, including extensions with opposite sign of exponential background, see the discussions and references in Ref.~\cite{son3}.
It is interesting to remark in this regard that the sign of $c$ in the exponential background was opposite already in the two pioneering papers~\cite{son2,andreev}.

An alternative to SW scenario for holographic description of linear Regge trajectories was proposed in Ref.~\cite{forkel}. The idea was to
introduce an infrared (IR) correction to the conformal dimension of operators, $\Delta\to\Delta+cz^2$, modeling the anomalous dimension.
Using the holographic relation between the 5D mass $m_5$ and $\Delta$, this is tantamount to introducing $\mathcal{O}(z^2)$ and $\mathcal{O}(z^4)$
IR modifications of $m_5^2$ in the holographic action. The $\mathcal{O}(z^4)$ and $\mathcal{O}(z^2)$ contributions generate the slope and intercept of
linear Regge trajectories, correspondingly. An important advantage was that the given approach could describe the linear Regge trajectories in light baryons,
while the usual SW model failed because the background can be factorized from the Dirac equation in the AdS space.
A very similar IR modification of $m_5^2$ arises after a special field redefinition
absorbing the SW background $e^{cz^2}$, this was first observed in Ref.~\cite{No-wall} and coined the ``No-wall'' holographic model.
Some important aspects of interrelation between the SW models with different sign of $c$ in the SW background $e^{cz^2}$ were analyzed in Ref.~\cite{gutsche}.
The transition from one form to the other was shown to be accompanied by the appearance of $z$-dependent mass terms. Various effective IR
modifications of 5D mass also emerge or are introduced in some other situations (see, e.g., the discussions and references in Ref.~\cite{Vega:2010ne}),
for instance, in describing the higher spin fields in the AdS space~\cite{br3,br2}.
In an extensive study of bottom-up holographic approach performed in Ref.~\cite{Gursoy:2007er}, the most general quadratic non-derivative term arising from fluctuations
of bulk fields was also a function of the fifth coordinate that is tantamount to a $z$-dependent mass term.
Another one predecessor of our present paper is Ref.~\cite{genSW}, where it was shown how to
introduce an arbitrary intercept in the linear trajectories via a certain generalization of the SW exponential background, the resulting
model remains closed-form solvable. Finally the present research was triggered by our recent work~\cite{AS} on a consistent holographic renormalization
of two-point correlators that leads to correct low-energy predictions.

The paper is organized as follows. The SW model for arbitrary integer spin is recalled in Section~2. Our holographic theory is introduced in Section~3.
In Section~4, we derive a generalized SW background for arbitrary integer spin. In Section~5, several applications of our approach are considered.
The obtained results are used in Section~5.1 for analyzing some confinement properties.
The discussions on embedding the effects related with the chiral symmetry breaking are given in Section~5.2. In Section~5.3, we discuss in detail the two-point vector correlator.
The pion form factor is studied in Section~5.4. Various technical details are shifted to the Appendices A, B, and C. The experimental phenomenology of
linear Regge trajectories is discussed in the Appendix~D. The Appendix~E is devoted to a brief review of the spectroscopy of Veneziano like dual amplitudes.

\section{Some preliminaries}

The action of SW model~\cite{son2} for free 5D scalar fields is
\begin{equation}
\label{31a}
S=\frac{1}{2}\int d^4\!x\,dz\sqrt{g}\,e^{cz^2}\left(\partial^M\Phi\partial_M\Phi-m_5^2\Phi^2\right).
\end{equation}
Here $g=|\text{det}g_{MN}|$ and a normalization constant for the 5D
fields $V_M$ will be omitted in what follows. The background space represents the Poincar\'{e} patch
of the AdS$_5$ space with the metric
\begin{equation}
\label{2}
g_{MN}dx^Mdx^N=\frac{R^2}{z^2}(\eta_{\mu\nu}dx^{\mu}dx^{\nu}-dz^2),\qquad z>0,
\end{equation}
where $\eta_{\mu\nu}=\text{diag}(1,-1,-1,-1)$, $R$ denotes the radius of AdS$_5$ space,
and $z$ is the holographic coordinate.
According to the standard prescriptions of AdS/CFT correspondence~\cite{witten,gub}
the 5D mass $m_5$ is determined by the behavior of 5D fields near the UV boundary $z=0$,
\begin{equation}
\label{3}
m_5^2R^2=\Delta(\Delta-4),
\end{equation}
where $\Delta$ stands for the scaling dimension of a 4D operator $\mathcal{O}(x_\mu)$ dual to the corresponding 5D field $\Phi(x_\mu,z)$.
The term ``dual'' means the identification~\cite{witten,gub}
(exhaustive discussions of this point are contained in Ref.~\cite{sundrum}),
\begin{equation}
\label{UVcond}
\left.\Phi(x_\mu,z)\right|_{z\rightarrow0}=z^\Delta\mathcal{O}(x_\mu).
\end{equation}
Sometimes one asks why should we use the AdS/CFT prescription~\eqref{3} in phenomenological bottom-up holographic models?
The matter is that~\eqref{3} is a consequence of the holographic identification~\eqref{UVcond}~\cite{sundrum}.
If we give the prescription~\eqref{3} up then an important connection with the gauge/gravity correspondence would be lost
and it would not be clear why we call our models ``holographic''. Also a deviation from~\eqref{3} results in a bad consequence
for the two-point correlation functions calculated using the AdS/CFT prescriptions: The leading logarithmic behavior disappears.
It should be recalled that the logarithmic asymptotic of QCD correlators emerges due to approximate scale invariance of
strong interactions at very high energies (the prescription~\eqref{3} is also rooted in the scale invariance). Consequently,
if we want to build a holographic model that interpolates QCD from low to high energies, we must impose the relation~\eqref{3},
at least in the UV limit $z\rightarrow0$. Some additional remarks on this issue are given after a generalization of~\eqref{3}
to higher spins, the relation~\eqref{3c}.

The spectrum of 4D modes of the model~\eqref{31a} is discrete and given by the relation~\eqref{primary} of Appendix~A for spin $J=0$,
\begin{equation}
\label{33a}
m_n^2=2|c|\left(2n+1+\sqrt{4+m_5^2R^2}-\frac{c}{|c|}\right),\qquad
n=0,1,2,\dots.
\end{equation}
This spectrum has a Regge form due to the dilaton background $e^{cz^2}$ in the 5D action~\eqref{31a}.

The construction of SW action~\eqref{31a} contains an explicit violation of scale and Lorentz invariance along the holographic coordinate $z$.
Anticipating our discussion in the next Section, we notice that the given setup represents only a minimal possibility. Actually if we wish
to describe a Regge-like spectrum of the kind~\eqref{33a} within a closed-form solvable SW holographic model, we may consider the following
general ansatz,
\begin{equation}
\label{31b}
S=\frac{1}{2}\int d^4\!x\,dz\sqrt{g}\,e^{cz^2}\left[\partial^M\Phi\partial_M\Phi-(m_5^2+a_1z^2+a_2z^4)\Phi^2+bg^{zz}\Phi z\partial_z\Phi\right],
\end{equation}
where the metric factor $g^{zz}$ originates from the covariant term \(g^{MN}\Phi x_M\partial_N\Phi\).
This extension of SW model will be analyzed below for arbitrary integer spin.

We recall briefly how the model~\eqref{31a} is generalized to arbitrary 5D tensor fields
$\Phi_J\doteq\Phi_{M_1M_2\dots M_J}$, $M_i=0,1,2,3,4$.
These fields describe higher spin mesons if the corresponding tensors are symmetric plus
some additional constraints are imposed. By assumption, the given tensor fields are dual
to QCD operators $\mathcal{O}_J\doteq\mathcal{O}_{\mu_1\mu_2\dots \mu_J}$, $\mu_i=0,1,2,3$,
near the boundary $z=0$.
The holographic duality entails the following generalization of the relation~\eqref{UVcond},
\begin{equation}
\label{UVcond2}
\left.\Phi_J(x_\mu,z)\right|_{z\rightarrow0}=z^{\Delta-J}\mathcal{O}(x_\mu),
\end{equation}
which generalizes the 5D mass~\eqref{3} (see, e.g., Ref~\cite{br3}),
\begin{equation}
\label{3c}
m_5^2R^2=(\Delta-J)(\Delta+J-4).
\end{equation}
It is important to emphasize that the internal self-consistency of the entire approach requires to
follow the prescription~\eqref{3c} because it guarantees the fulfilment of the duality condition~\eqref{UVcond2}.
The constant part of a 5D mass, consequently, cannot be taken arbitrary. The 5D mass of gauge higher spin
fields in the original SW model~\cite{son2} does not satisfy~\eqref{3c} since the condition of
generalized gauge invariance in the AdS$_5$ space is incompatible with the prescription~\eqref{3c},
at least when the dimensions of quark fields or of derivatives of gluon fields contribute to $\Delta$.
We will follow~\eqref{3c}, some additional arguments are given in Ref.~\cite{br2}.

Since a physical hadron has polarization indices along the usual 3+1 physical coordinates,
a projection to 4D particle states in holographic QCD is usually achieved via the condition,
\begin{equation}
\label{18}
\Phi_{z\dots}=0,
\end{equation}
i.e. the physical components are $\Phi_{\mu_1\mu_2\dots \mu_J}$, all other components vanish identically.
A generalization of action~\eqref{31a} to the case of arbitrary integer spin reads
\begin{equation}
\label{19}
S=\frac12\int d^4x\,dz \sqrt{g}\,e^{cz^2}\!\left(D^M\Phi^J D_M\Phi_J-m_5^2\Phi^J\Phi_J\right).
\end{equation}
Two comments are in order. First, an action for free higher spin fields contains many quadratic terms appearing from many ways of
contraction of coordinate indices. But the condition~\eqref{18} greatly simplifies the action leaving only two terms explicitly
shown in~\eqref{19}~\cite{son2,br3,br2}. Second, the effect of AdS affine connections in the covariant derivatives $D_M$ leads
just to a constant shift of the mass term. In the presence of $z$-dependent dilaton background, this shift becomes also $z$-dependent.
In the case of quadratic dilaton of the action~\eqref{19}, the shift has a structure $m_5^2\rightarrow c_1z^2+c_2$, where $c_1$ and $c_2$
are certain constants~\cite{br2}. As the given structure represents a particular case of more general ansatz~\eqref{31b} we will consider,
the covariant derivatives $D_M$ in~\eqref{19} can be replaced by normal derivatives.

The spectrum of the model~\eqref{19} has the Regge form in both the spin $J$ and radial $n$ directions but details of the spectrum depend on
additional assumptions~\cite{son2,br3,gutsche,afonin2010}. If the 5D higher spin fields are treated as massive ones with masses
dictated by~\eqref{3c} and imposed constraint~\eqref{18}, the spectrum of physical modes is determined from the equation of motion
\begin{equation}
\label{eom}
\frac{\delta S}{\delta\Phi_{\mu_1\mu_2\dots \mu_J}}=0,
\end{equation}
while the Euler-Lagrange equations
\begin{equation}
\frac{\delta S}{\delta\Phi_{zM_2\dots M_J}}=0,
\end{equation}
will give the kinematical constraints $\partial^{\mu}\Phi_{\mu\dots}=0$ and $\eta^{\mu\nu}\Phi_{\mu\nu\mu_3\dots\mu_J}=0$
eliminating the lower-spin states and thus providing the required $2J+1$ physical degrees of freedom~\cite{br3,br2}.
The further Euler-Lagrange equations will not appear due to the constraint~\eqref{18}.

The equation of motion~\eqref{eom} together with the condition~\eqref{3c}
leads to the spectrum (see Appendix~A),
\begin{equation}
\label{33b}
m_{n,J}^2=2|c|\left(2n+\Delta-1+\frac{c}{|c|}(J-1)\right).
\end{equation}
For the most interesting case of twist-2 operators\footnote{In the theory of deep inelastic scattering, the twist corresponds to the number of parton
constituents inside a hadron. Since a meson consists of a quark-antiquark pair and the higher Fock components are suppressed in the large-$N_c$
limit of QCD~\cite{hoof}, the case of twist-2 seems to be the most relevant to the holographic approach.}, $\Delta=J+2$, we obtain
\begin{equation}
\label{33c}
m_{n,J}^2=2|c|\left(2n+J+1+\frac{c}{|c|}(J-1)\right),\qquad
J>0.
\end{equation}
This spectrum has the string form for $c>0$,
\begin{equation}
\label{string}
m_{n,J}^2=4|c|\left(n+J\right),
\end{equation}
and formally does not depend on the spin $J$ for $c<0$.

\section{``Maximally extended'' closed-form solvable SW model}

Following the discussions of the previous Section, let us add to the SW action~\eqref{19} (with covariant derivatives replaced by the normal ones)
the $z$-dependent terms shown for the scalar case in the action~\eqref{31b},
\begin{equation}
\label{34}
  \begin{aligned}
    S&=\frac{1}{2}\int d^5x\sqrt{g}e^{cz^2}g^{M_1N_1}\dots g^{M_JN_J}\lsb
    g^{MN}\partial_M \Phi_{M_1\dots M_J}\partial_N\Phi_{N_1\dots N_J}-\right.\\&\left.
    \lb m_5^2 + a_1z^2+a_2z^4\rb \Phi_{M_1\dots M_J}\Phi_{N_1\dots N_J}+
    b g^{zz}\Phi_{M_1\dots M_J}z\pz\Phi_{N_1\dots N_J}\rsb.
  \end{aligned}
\end{equation}
To calculate the spectrum and find its parametric dependence it is convenient to eliminate the exponential dilaton
factor via the field replacement,
\begin{equation}
\label{35}
\Phi_{M_1\dots M_J} = e^{-cz^2/2}\phi_{M_1\dots M_J}.
\end{equation}
Substituting~\eqref{35} to~\eqref{34} we get
\begin{equation}
  \begin{aligned}
    S&=\frac{1}{2}\int d^5x\sqrt{g}g^{M_1N_1}\dots g^{M_JN_J}\lsb
    g^{\mu\nu}\partial_\mu\phi_{M_1\dots M_J}\partial_\nu\phi_{N_1\dots N_J}+\right.\\&
    g^{zz}\lb c^2z^2\phi_{M_1\dots M_J}\phi_{N_1\dots N_J}-2cz\phi_{N_1\dots N_J}\pz\phi_{M_1\dots M_J}+\pz\phi_{M_1\dots M_J}\pz\phi_{N_1\dots N_J}\rb+\\&
    \lb m_5^2 + a_1z^2+a_2z^4\rb \phi_{M_1\dots M_J}\phi_{N_1\dots N_J}+
    \left. bg^{zz}\phi_{M_1\dots M_J}z\lb\pz\phi_{N_1\dots N_J}-cz\phi_{N_1\dots N_J}\rb\rsb,
  \end{aligned}
\end{equation}
or, combining the similar terms,
\begin{align}
  S=\frac{1}{2}\int d^5x&\sqrt{g}g^{M_1N_1}\dots g^{M_JN_J}\lsb
  g^{MN}\partial_M\phi_{M_1\dots M_J}\partial_N\phi_{N_1\dots N_J}-\right.\nonumber\\&-\left.
  \lb 2cz-b z\rb g^{zz} \phi_{N_1\dots N_J}\pz\phi_{M_1\dots M_J}-\right.\\&-\left.
  \lb m_5^2 + a_1z^2 + a_2z^4-g^{zz}c^2z^2+b g^{zz}cz^2\rb\phi_{M_1\dots M_J}\phi_{N_1\dots N_J}\rsb.\nonumber
\end{align}
Using the AdS metric factor~\eqref{2} we obtain the final form before variation,
\begin{equation}
\label{36}
  \begin{aligned}
    S=\frac{1}{2}\int &d^5x\lb\frac{z^2}{R^2}\rb^{J-3/2}\lsb
    \left(\partial_M\phi_{\dots}\right)^2+
    z\lb 2c-b \rb\phi_{\dots}\pz\phi_{\dots}-\right.\\&-\left.
    \lb \frac{m_5^2R^2}{z^2}+a_1R^2+z^2\left(a_2R^2+c^2-bc\right)\rb\phi_{\dots}^2\rsb.
  \end{aligned}
\end{equation}
Here and below the lower contraction $\phi_{\dots}\phi_{\dots}$ means
the usage of flat metric \(\eta^{MN}\equiv\text{diag}\{1,-1,-1,-1,-1\}\)
when contracting lower and upper indices.

The variation of action~\eqref{36}, after integrating by parts and some simple algebra,
leads to the expression,
\begin{equation}
\label{37}
  \begin{aligned}
    \frac{\delta S}{\delta\phi_{\dots}}=\int d^5x\lsb
    -\partial_M\lb\lb\frac{z^2}{R^2}\rb^{J-3/2}\partial_M\phi_{\dots}\rb-
    \lb\frac{z^2}{R^2}\rb^{J-5/2}m_\text{eff}^2(z)\phi_{\dots}\rsb,
  \end{aligned}
\end{equation}
where the $z$-dependent effective mass is
\begin{equation}
m_\text{eff}^2(z)= m_5^2 + z^2\left(a_1+\frac{(J-1)(2c-b)}{R^2}\right) + z^4\left(a_2+\frac{c^2-bc}{R^2}\right).
\end{equation}
After obvious redefinitions, the effective 5D mass takes the form
\begin{equation}
\label{meff}
m_\text{eff}^2R^2 = m_5^2R^2 + \tilde{b}z^2+\tilde{c}^2z^4.
\end{equation}
It is seen now that although the original extended SW action~\eqref{34} contained 4 parameters, the spectrum
will effectively depend on 2 parameters.

The equation of motion stemming from~\eqref{37} is
\begin{equation}
\label{38}
\partial_\mu^2\phi_{\dots}-\lb\frac{R}{z}\rb^{2J-3}\partial_z\left(\lb\frac{z}{R}\rb^{2J-3}\partial_z\phi_{\dots}\right)+
\frac{m_\text{eff}^2R^2}{z^2}\phi_{\dots}=0.
\end{equation}
In the AdS/QCD models, the 4D particles are identified with normalizable solutions in the form of 4D plane waves,
\begin{equation}
\label{39}
\phi_{\mu_1\dots\mu_J}(x_\mu,z)=e^{ip_\mu x_\mu}v^{(J)}(z)\epsilon_{\mu_1\dots\mu_J},
\end{equation}
where $\epsilon_{\mu_1\dots\mu_J}$ denotes the polarization tensor and $v^{(J)}(z)$ is the profile function of physical 4D modes having the
mass squared $m^2=p_\mu^2$.
By the standard substitution,
\begin{equation}
\label{39b}
v^{(J)}=\lb\frac{z}{R}\rb^{3/2-J}\psi^{(J)},
\end{equation}
the Eq.~\eqref{39} is converted into one-dimensional Schr\"{o}dinger equation
\begin{equation}
-\partial_z^2\psi^{(J)}+V(z)\psi^{(J)}=m^2\psi^{(J)},
\end{equation}
with the potential of harmonic oscillator
\begin{equation}
\label{21}
V(z)=\tilde{c}^2z^2+\tilde{b}+\frac{(J-2)^2+m_5^2R^2-1/4}{z^2}.
\end{equation}
The normalizable spectrum of discrete modes, $n=0,1,2,\dots$, is
\begin{equation}
\label{40}
m_{n,J}^2=2|\tilde{c}|\left(2n+1+\sqrt{(J-2)^2+m_5^2R^2}+\frac{\tilde{b}}{2|\tilde{c}|}\right).
\end{equation}
Using as before the condition~\eqref{3c} for the twist-2 dimension $\Delta=J+2$, the spectrum
simplifies to
\begin{equation}
\label{41}
m_{n,J}^2=2|\tilde{c}|\left(2n+1+J+\frac{\tilde{b}}{2|\tilde{c}|}\right).
\end{equation}
Coming back to our original notations in the action~\eqref{34}, the parameters are
\begin{equation}
\label{42}
\tilde{b}=a_1R^2+(2c-b)(J-1),\qquad \tilde{c}=a_2R^2+c(c-b).
\end{equation}

The spectrum has the Regge form and as it follows from the derivation, the considered ansatz
is the most general one among closed-form solvable cases leading to exactly linear Regge trajectories.
In the literature, one can find some closed-form solvable extensions of SW model possessing an arbitrary intercept~\cite{genSW,forkel,AS}.
The spectrum~\eqref{41} is more general --- the slopes of the radial and spin trajectories can be in arbitrary proportion.
This proportion is regulated by the parameter $b$, i.e., by the term linear in derivative in the action~\eqref{34}.
In some phenomenological analyses of light meson spectra, these slopes turn out to be different indeed, as in Ref.~\cite{arriola}.
Our holographic theory thus can easily accommodate this scenario while for the standard SW holographic models this would be problematic.

The standard SW holographic model is known to work differently for different sign of mass parameter $c$ in the dilaton background $e^{cz^2}$.
Various aspects of this difference were discussed in the past~\cite{br3,SWaxial,afonin2010,son3,zuo,gutsche}. A particular manifestation
is the mentioned independence of spin in the spectrum~\eqref{33c} if $c<0$. It is instructive to see how the transition from $c>0$ to $c<0$
happens within our general case. For the sake of definiteness, imagine that we want to obtain the spectrum of Veneziano dual amplitude
(see the relation~\eqref{venSP} in the Appendix~E),
\begin{equation}
\label{43}
m_{n,J}^2=4\lambda^2\left(n+J\right),
\end{equation}
that was reproduced in the first SW holographic model~\cite{son2}. As follows from~\eqref{41} and~\eqref{42},
in the case of positive sign, $c>0$, this spectrum is reproduced if we set $a_1=a_2=b=0$, $c=\lambda^2$. In the case of negative dilaton background, $c<0$,
the solution becomes more complicated: $a_1=0$, $b=4c$, $a_2R^2=4c^2$, $c=-\lambda^2$. The transition from the model with positive sign to the model
with negative one can be achieved via the field replacement,
\begin{equation}
\label{44}
\Phi_{M_1\dots M_J}^+ = e^{-cz^2}\Phi_{M_1\dots M_J}^-.
\end{equation}
A similar replacement was exploited in Ref.~\cite{gutsche}. Our approach yields automatically the form and magnitude of additional terms which arise in the 5D action
after such a replacement. It should be also noted that if the higher spin fields are described as pure gauge 5D fields, when the relation~\eqref{3c}
near the UV boundary is not imposed, the situation is opposite --- the simplest solution, $a_1=a_2=b=0$, $c=-\lambda^2$, is achieved when $c<0$.
Exactly this description was exploited in the original paper where the SW holographic model was introduced~\cite{son2}.

Concluding the discussions of this section, it is worth mentioning the first AdS/QCD model where a $z$-dependent IR modification of 5D mass was proposed~\cite{forkel}.
The idea of Ref.~\cite{forkel}, in a somewhat reformulated version, was to introduce a $z$-dependent IR modification for conformal dimension $\Delta$ in
the relation~\eqref{3c}. The suggested ansatz is
\begin{equation}
\Delta\rightarrow\Delta +cz^2,
\end{equation}
that after substitution to~\eqref{3c} results in the following IR modification of 5D mass,
\begin{equation}
\label{meffFor}
m_5^2R^2\rightarrow m_5^2R^2+2Jcz^2 +c^2z^4,
\end{equation}
where we set $\Delta-2=J$ for the twist-two operators. The $z$-dependent 5D mass~\eqref{meffFor} is a particular case of our general ansatz~\eqref{meff}.
The Regge spectrum of this model follows immediately from~\eqref{41}, where we must substitute $\tilde{c}=c$ and $\tilde{b}=2cJ$.
For the physical case $c>0$ we obtain
\begin{equation}
\label{41For}
m_{n,J}^2=4c\left(n+J+\frac12\right).
\end{equation}
It was also suggested to replace the total spin $J$ in the Regge spectrum~\eqref{41For} by the orbital momentum $L$ of quark-antiquark pair.
The resulting spectrum becomes then very similar to the spectra~\eqref{cluster} or~\eqref{LplusN} in the Appendix D, which were extracted from
the experimental data. It is not surprising that the agreement with the data presented in Ref.~\cite{forkel} looked impressive.

\section{Relation between the SW background and $z$-dependent 5D mass}

Our previous analysis demonstrated that a closed-form solvable SW model with linear spectrum allows for 5D mass to have two $z$-dependent
contributions as is shown, for example, in the relation~\eqref{meff}. In the practical situations, the $\mathcal{O}(z^4)$ contribution
can be absorbed by redefinition of the mass parameter in the SW exponential background. The given term contributes to the slope of Regge trajectories.
The $\mathcal{O}(z^2)$ contribution dictates the intercept. This contribution is very important as it determines the mass of
the ground state and also may describe the mass shifts caused by the spontaneous chiral symmetry breaking. A question appears whether an arbitrary
intercept parameter can be also parametrized by some correction to the SW background? A solution for such a background was found in Ref.~\cite{genSW}.
It was erroneously concluded, however, that this background is not fully equivalent to the $\mathcal{O}(z^2)$ contribution to 5D mass term and that
the equivalence appears in the infrared limit, $z\rightarrow\infty$, only . We are going to correct this error and show by an explicit calculation that
the two forms of SW model in question are fully equivalent. We will begin with the vector case which was analyzed in the original paper~\cite{genSW}.
After that we generalize the result to the tensor case and find the appropriate background for any integer spin.

\subsection{A simple vector model}

Consider the following problem: How to reproduce the linear Regge spectrum of vector mesons,
\begin{equation}
\label{vect}
m_n^2=4c\left(n+1+b\right),\qquad n=0,1,2,\dots,
\end{equation}
with $c>0$ and arbitrary intercept parameter $b$, within a SW holographic model with massless 5D vector field?
The solution derived in Ref.~\cite{genSW} reads
\begin{equation}
\label{45b}
S=\Gamma^2(1+b)\int d^5 x \sqrt{g}e^{-cz^2}U^2(b,0,cz^2)g^{MR}g^{NS}\left(-\frac14 F_{MN}F_{RS}\right),
\end{equation}
where $U$ is the Tricomi confluent hypergeometric function. The case of $c<0$ was not considered in Ref.~\cite{genSW} and will
be discussed later. Below we give a short proof of the statement that the action~\eqref{45b} is fully equivalent to
writing a $\mathcal{O}(bz^2)$ mass term in a theory without the Tricomi function in the background.
A more rigorous proof is presented in the Appendix~B.

Using the axial gauge~\eqref{18} and omitting the Lorentz indices, the term in~\eqref{45b} relevant to the problem
under consideration is (we set $R=1$)
\begin{equation}
\label{45c}
S_{(z)} \sim  \int dz\,\frac{e^{-cz^2}}{z}\,U^2(b,0,cz^2)\left(\partial_zV\right)^2.
\end{equation}
Let us eliminate the nontrivial background with the help of the field redefinition
\begin{equation}
  V=\frac{e^{cz^2/2}}{U}\,v.
\end{equation}
We do not write the arguments of the Tricomi function $U(b,0,cz^2)$ here and in what follows.
The result is
\begin{multline}
S_{(z)} \sim  \int \frac{dz}{z} \left[\left(\partial_zv\right)^2+c^2z^2v^2+\left(\frac{\partial_z U}{U}\right)^2v^2 + \right. \\
\left. 2\left(cz-\frac{\partial_z U}{U}\right)v\partial_zv-2cz\frac{\partial_z U}{U}v^2\right].
\label{45d}
\end{multline}
The equation of motion stemming from~\eqref{45d} is
\begin{equation}
\label{eom45d}
\partial_z\left(\frac{\partial_z v}{z}\right)-\left[\frac{1}{z}\left(cz-\frac{\partial_z U}{U}\right)^2 + \partial_z\left(\frac{\partial_z U}{zU}\right) \right]v=0.
\end{equation}
Using the derivatives of Tricomi function,
\begin{equation}
\label{derU}
\partial_z U(b,0,cz^2)=2czU', \qquad \partial_z^2 U(b,0,cz^2) = 2cU' + 4c^2z^2U'',
\end{equation}
where $U'(b,k,x)\doteq\partial_x U(b,k,x)$, the equation~\eqref{eom45d} takes the form
\begin{equation}
\label{2eom45d}
\partial_z\left(\frac{\partial_z v}{z}\right)-\frac{c^2}{z}\left[z^2+4z^2\frac{U''-U}{U}\right]v=0.
\end{equation}
The Tricomi function $\omega=U(b,k,x)$ is a solution of Kummer's equation~\eqref{kummer} (see the Appendix~C for the related details),
\begin{equation}
\label{kum}
x\omega''+(k-x)\omega'=b\omega.
\end{equation}
In our case of $k=0$, the Eq.~\eqref{kum} yields $x(U''-U')=bU$, that transforms the Eq.~\eqref{2eom45d} into
\begin{equation}
\label{3eom45d}
\partial_z\left(\frac{\partial_z v}{z}\right)-\frac{1}{z}\left[c^2z^2+4cb\right]v=0.
\end{equation}
After the transition to the form of a Schr\"{o}dinger equation, the term with $b$ becomes the constant
contribution in the potential~\eqref{21} which arises from the $\mathcal{O}(bz^2)$ contribution to the effective 5D mass~\eqref{meff}.
This proves our statement.

Consider now the case $c<0$. Repeating the derivation above it is seen that for arbitrary sign of $c$ the equation~\eqref{2eom45d}
is replaced by
\begin{equation}
\label{4eom45d}
\partial_z\left(\frac{\partial_z v}{z}\right)-\frac{c^2}{z}\left[z^2+4z^2\frac{U''-\frac{c}{|c|}U}{U}\right]v=0.
\end{equation}
For $c<0$, the Kummer's equation~\eqref{kum} can also be used if we replace $x$ by $(-x)$.
The solution formally becomes
\begin{equation}
\label{cs}
c<0:\qquad \omega=U(-b,0,-|c|z^2).
\end{equation}
The Tricomi function $U(b,k,x)$, however, is complex
for integer $k$ when its argument is negative, $x<0$. The real-valued solution formally is $\omega=M(-b,0,-|c|z^2)$,
where $M$ denotes the Kummer function, the second solution of Kummer's equation (see the Appendix~C).
But the function $M(b,k,x)$ is not defined when $k$ an integer less than 1, thus it is also problematic
in our case $k=0$. A well defined solution is given by (see the Appendix~C)
\begin{equation}
\label{solo}
\omega=|c|z^2M(1-b,2,-|c|z^2).
\end{equation}
Unfortunately, this solution is not acceptable for our purpose: since
\begin{equation}
M(b,k,x)|_{x\rightarrow0}=1+\mathcal{O}(x),
\end{equation}
we do not obtain the condition $\omega(0)\sim 1$ in the action~\eqref{45b} (where $\omega=U$). The given restriction provides the
necessary asymptotics of AdS$_5$ space near the boundary.
In addition, the infrared asymptotics of the function~\eqref{solo},
\begin{equation}
\omega_{z\rightarrow\infty}\sim e^{-|c|z^2}z^{-2b},
\end{equation}
would forbid the positive exponent in the background of action~\eqref{45b}: As the background is proportional to $e^{-cz^2}\omega^2$,
we would obtain $e^{-|c|z^2}$ at $c<0$ and large $z$.
We thus arrive at the conclusion that the case $c<0$ does not work in the given situation.

One may try to cope with the complexity of solution~\eqref{cs} replacing the factor $U^2$ in the background of action~\eqref{45b}
by $UU^*$ and writing a model for charged vector fields, this would replace the term~\eqref{45c} by
\begin{equation}
\label{45e}
S_{(z)} \sim  \int dz\,\frac{e^{-cz^2}}{z}\,U(-b,0,-cz^2)U^{*}(-b,0,-cz^2)\partial_zV\partial_zV^*.
\end{equation}
But one can show that such a formulation does not work as well.

\subsection{Generalized SW model for tensor fields}

Now we are ready to derive a generalization of background in the SW action~\eqref{45} to the case of arbitrary integer spin.
Consider a free action for massless tensor field in flat 5D space,
\begin{equation}
\label{genT}
  S=\frac{1}{2}\int d^5x f^2(z) \lb\partial_M \Phi_{M_1\dots M_J}\rb^2,
\end{equation}
where \(f(z)\) is a background function which we intend to find in this section. Following~\cite{genSW}
we hide inside \(f^2\) other background-related factors such as
\(\sqrt{g}\) and metrics that are used to contract indices. The constant mass term can be easily added at the end
and will not change the result.
In order to get the AdS rules for contraction of Lorentz indices we must impose the following UV asymptotics,
\begin{equation}
  f(z)\underset{z\to0}{\sim} z^{(2J-3)/2}.
\end{equation}
With this asymptotics, the action~\eqref{genT} can be formally written in the form of action~\eqref{34}.

With the condition~\eqref{18}, the dynamical equation of motion reads
\begin{equation}
  \lsb f^2(z)\partial_\mu\partial^\mu-
  \pz\lb f^2(z)\pz\rb\rsb\Phi^{\mu_1\dots \mu_J}=0,
\end{equation}
where the indices are contracted using the flat metric
\begin{equation}
  \eta^{MN}\equiv\text{diag}\{1,-1,-1,-1,-1\}.
\end{equation}
Using the notations~\eqref{39}, the equation takes the following form,
\begin{equation}
  \lsb f^2m^2+\pz\lb f^2\pz\rb\rsb v^{(J)}=0.
\end{equation}
After the substitution
\begin{equation}
v^{(J)}=\frac{\psi^{(J)}}{f},
\end{equation}
we get the Schr\"{o}dinger equation
\begin{equation}
\label{eqf}
  -\partial^2_z\psi^{(J)}+\frac{\partial^2_zf}{f}\psi^{(J)}=m^2\psi^{(J)}.
\end{equation}

We wish to obtain the Regge spectrum of massless tensor fields in AdS$_5$ space in the form of~\eqref{primary}
(for $m_5=0$) but now shifted by the intercept parameter $b$, $n\rightarrow n+b$, i.e.
\begin{equation}
\label{vect2}
  m_n^2=2|c|\lb 2n+1+|J-2|+\frac{c}{|c|}(J-1)+2b\rb.
\end{equation}
According to Appendix A, the potential then must be
\begin{equation}
\label{diff}
  \frac{\partial^2_zf}{f}=c^2z^2+2c(J-1)+\frac{(J-2)^2-1/4}{z^2}+4b|c|.
\end{equation}
This second-order differential equation for $f(z)$ has two linearly independent solutions
in terms of confluent hypergeometric functions.
As is well known from the theory of such equations,
a linear combination of these solutions can be chosen in such a way that one of them is exponentially
decreasing in the (\(z\to\infty\)) asymptotics, the second one will be exponentially growing.
Imposing the normalization condition (to preserve the standard normalization factor in the action~\eqref{genT})
\begin{equation}
  f^2(z)z^{3-2J}\eqzz 1,
\end{equation}
these solutions can be written as follows,
\begin{multline}
\label{sol1b}
  f_1(z)=\frac{\Gamma\lb2-J+(J-1)\frac{c+|c|}{2|c|}+b\rb}{\Gamma(2-J)} e^{-|c|z^2/2} z^{(2J-3)/2}\times\\\times
  U\lb\frac{J-1}{2}\lb1+\frac{c}{|c|}\rb+b,J-1,|c|z^2\rb,
\end{multline}
\begin{equation}
\label{sol2b}
  f_2(z)=e^{|c|z^2/2} z^{(2J-3)/2}{}M\lb\frac{J-1}{2}\lb1-\frac{c}{|c|}\rb-b,J-1,-|c|z^2\rb,
\end{equation}
where $U$ is the Tricomi function and $M$ is the Kummer function (which represents the confluent hypergeometric function $_1F_1$).
The latter can be expressed via the Laguerre function (the generalization to fractional powers of argument of associated
Laguerre polynomials) which appear as eigenfunctions of normalizable solutions, see Eq.~\eqref{wf}, using the relation:
$L^{k}_{b}(x)=\binom{b+k}b\,_1F_1\left(-b,k+1,x\right)$.
These solutions have different asymptotics. The first solution converges both at \(z\to0\)
and \(z\to\infty\), while the second one converges at \(z\to0\) and diverges at \(z\to\infty\)
due to $U(b,\alpha,x)\sim L_{-b}^\beta(x) \sim x^{-b}$ at
\(x\to\infty\) for arbitrary $\alpha$ and $\beta$, related as $\alpha=\beta+1$ in the case under consideration.

There are various subtle points with the solution of equation~\eqref{diff} because of which
the results obtained from popular software systems like Mathematica should be used with care.
Ignoring this can lead to (partly) incorrect statements which can be met in some papers.
In the Appendix~C, we find the solutions~\eqref{sol1b} and~\eqref{sol2b} analytically and discuss the arising
subtleties.

Now we should interpret~\eqref{sol1b} and~\eqref{sol2b} as generalization of SW model with positive and negative dilaton background, SW$^\pm$ for brevity.
The standard SW$^\pm$ model must appear at zero intercept parameter $b=0$. Since $U(0,\alpha,x)=M(0,\alpha,x)=1$ for any $\alpha$,
it looks natural to choose $c<0$ in the first solution~\eqref{sol1b} and interpret it as an extension of SW$^-$ model.
After that we cannot set $c<0$ in the second solution~\eqref{sol2b} because $M(\alpha,\alpha,x)=e^x$, hence, we would obtain the same SW$^-$ background.
Thus we must choose  $c>0$ in~\eqref{sol2b}. The solutions
\begin{equation}
\label{sol1}
  f_1(z)=\frac{\Gamma\lb2-J+b\rb}{\Gamma(2-J)} e^{-|c|z^2/2} z^{(2J-3)/2} U\lb b,J-1,|c|z^2\rb,
\end{equation}
\begin{equation}
\label{sol2}
  f_2(z)=e^{|c|z^2/2} z^{(2J-3)/2}{}M\lb-b,J-1,-|c|z^2\rb,
\end{equation}
would define then the backgrounds $f_1^2(z)$ and $f_2^2(z)$ of generalized SW$^-$ and SW$^+$ models, correspondingly.
However, the Kummer function $M(b,\alpha,x)$ is not a solution (is not defined as some entire function of $x$) if $\alpha$ is an integer less than 1.
This excludes the cases $J=0$ and $J=1$. To include the scalar and vector cases we can use a linearly independent solution
or define a certain linear combination of them called Tricomi function $U$ (see the Appendix~C).
That was the reason for writing the first solution~\eqref{sol1b} in terms of $U$, in this form it suits for any spin.
The function $U(b,\alpha,x)$ is formally undefined for integer $\alpha$
but can be analytically extended to any integer $\alpha$ by continuity. This extension, however,
results in complex values at $x<0$, so we cannot use it in the second solution~\eqref{sol2b}.
It should be mentioned also that if we wrote $M$ instead of $U$ in~\eqref{sol1b} and chose the same sign of $c$ in both solutions,
then the solutions~\eqref{sol1b} and~\eqref{sol2b} would not be independent, they were related in this case via
the Kummer's transformation $M(b,\alpha,x) = e^x M(\alpha-b,\alpha,-x)$.

Further we notice that at $c<0$ the mass spectrum~\eqref{vect2} does not depend on spin for tensor mesons.
The choice of $c<0$ for $J>1$ is thus unphysical in our approach (but not necessarily if tensor fields are introduced in a different way).
Combining this observation with the discussion above, we
arrive at the conclusion that a self-consistent extension of the exponential background of the $SW$ model that produces an arbitrary intercept $b$
exists only in the following cases:
For SW$^-$ if $J=0,1$ and for SW$^+$ if $J>1$. The corresponding SW backgrounds are $f^-(z)=f_1^2(z)$ and $f^+(z)=f_2^2(z)$, where $f_1(z)$
and $f_2(z)$ are given by the relations~\eqref{sol1} and~\eqref{sol2}.

Substituting the solutions~\eqref{sol1} and~\eqref{sol2} into the action~\eqref{genT},
moving the \(z^{2J-3}\) factor back into \(\sqrt{g}\) and in the AdS contraction rule for Lorentz indices, we finally get
the extensions of SW holographic model to arbitrary intercept $b$,
\begin{equation}
\label{genJ}
  S_{J\leq1}=\frac{\Gamma^2\lb 2-J+b\rb}{2\Gamma^2(2-J)}\int d^5x\sqrt{g}e^{-|c|z^2}U^2\lb b,J-1,|c|z^2\rb
  \partial^M \Phi^{(J)}\partial_M\Phi^{(J)},
\end{equation}
for $J=0,1$ (for $J=1$, it reduces to the action~\eqref{45} in the Appendix~B) and
\begin{equation}
\label{genJ2}
  S_{J>1}=\frac{1}{2}\int d^5x\sqrt{g}e^{|c|z^2}M^2\lb -b,J-1,-|c|z^2\rb
  \partial^M \Phi^{M_1\dots M_J}\partial_M\Phi_{M_1\dots M_J},
\end{equation}
for $J>1$.

The generalized background with the Kummer function can be rewritten as the standard SW background
following the same steps as in the previous section for the vector case, this also results in the
appearance of $O(z^2)$ contribution to the 5D mass.

It is not difficult to show that the inclusion of constant mass does not change the result.
Indeed, in the axial gauge~\eqref{18}, the action~\eqref{genT} becomes
\begin{equation}
  S=\frac{1}{2}\int d^5x f^2(z) \lsb\lb\partial_M \Phi_{M_1\dots M_J}\rb^2-
  \frac{m_5^2R^2}{z^2}\lb\Phi_{M_1\dots M_J}\rb^2\rsb.
\end{equation}
Repeating the same steps as above, we obtain the massive extension of Eq.~\eqref{eqf}
\begin{equation}
\label{eqf2}
  -\partial^2_z\psi^{(J)}+\lb\frac{\partial^2_zf}{f}+\frac{m_5^2R^2}{z^2}\rb\psi^{(J)}=m^2\psi^{(J)},
\end{equation}
and require that its discrete spectrum must be given by the ``massive'' extension of spectrum~\eqref{vect2},
\begin{equation}
\label{vect3}
  m_n^2=2|c|\lb 2n+1+\sqrt{(J-2)^2+m_5^2R^2}+\frac{c}{|c|}(J-1)+2b\rb.
\end{equation}
This leads to the condition
\begin{equation}
\label{diff2}
  \frac{\partial^2_zf}{f}+\frac{m_5^2R^2}{z^2}=c^2z^2+2c(J-1)+\frac{(J-2)^2+m_5^2R^2-1/4}{z^2}+4b|c|.
\end{equation}
The terms with $m_5^2$ cancel and we arrive at the condition~\eqref{diff}.

\subsection{Brief summary}

Let us briefly summarize the main points of Sections~3 and~4. \\
(i) The general structure of closed-form solvable
SW holographic model describing the linear Regge like spectrum of spin-$J$ mesons is 
\begin{equation}
S=\frac{1}{2}\int d^4\!x\,dz\sqrt{g}\,e^{\pm|c|z^2}f^2(b,J,z)\left[\partial^M\Phi_J\partial_M\Phi_J-m_5^2(b,J,z)\Phi_J^2\right],
\end{equation}
where $\Phi_J$ denotes a spin-$J$ field (Lorentz indices are omitted), $\pm$ refers to the SW$^{\pm}$ variant of the model,
$f^2(b,J,z)$ is a background function introducing the intercept parameter $b$, and $m_5^2(b,J,z)$ is a mass function that also can introduce
the intercept $b$. \\
(ii) The dependence on $b$ can be completely hidden in $f(b,J,z)$ only in two cases: (1) $J=0,1$ in the SW$^{-}$ model;
$J>1$ in the SW$^{+}$ model. The mass function is then a constant, $m_5^2(b,J,z)=m_5^2(J)$, where $m_5^2(J)$ is dictated by the usual AdS/CFT 
prescription for 5D mass; the background function $f(b,J,z)$ for both cases was found in Section~4.2. \\
(iii) In all cases of SW$^{\pm}$ model, however, the intercept parameter $b$
can be introduced via the following phenomenological $z$-dependence of mass function (in units of $R=1$),
\begin{equation}
\text{SW}^\pm:\qquad m_5^2(b,J,z)=m_5^2(J)+b|c|z^2,
\end{equation}
and with trivial background function $f(b,J,z)=1$.\\
(iv) In the No-wall variant of the model, i.e., when the whole background is trivial, $e^{\pm|c|z^2}f^2(b,J,z)=1$, the mass function must
introduce also the slope, the corresponding ansatz is
\begin{equation}
\text{SW}^0:\qquad m_5^2(b,J,z)=m_5^2(J)+b|c|z^2+c^2z^4.
\end{equation}

\section{Some applications}

\subsection{Warped metrics and confining behavior}

Consider again the action for a 5D spin-$J$ field $\Phi_J$
\begin{equation}
\label{con1}
S=\frac12\int d^4x\,dz \sqrt{g}\,B(z)\!\left(D^M\Phi^J D_M\Phi_J-m_5^2\Phi^J\Phi_J\right),
\end{equation}
with the metrics (see~\eqref{2})
\begin{equation}
g_{MN}=\frac{R^2}{z^2}\eta_{MN},
\end{equation}
and some $z$-dependent background $B(z)$.
It is straightforward to see that this action can be formally rewritten as
\begin{equation}
\label{con2}
S=\frac12\int d^4x\,dz \sqrt{\tilde{g}}\left(D^M\Phi^J D_M\Phi_J-m_5^2B^{-(3/2-J)^{-1}}\Phi^J\Phi_J\right),
\end{equation}
with the warped metrics
\begin{equation}
\label{warped}
\tilde{g}_{MN}=B^{(3/2-J)^{-1}}g_{MN}.
\end{equation}
In principle, one may speculate that the action~\eqref{con2} defines a holographic model with certain $z$-dependent
mass term (or vice versa one can impose a $z$-dependence on $m_5^2$ in~\eqref{con1} in such a way that the mass term in~\eqref{con2} is constant)
and that this model can take the form of action~\eqref{con1} if one neglects the affine connections in covariant derivatives.

What can we gain in hiding the $z$-dependence into the metrics? A real advantage for the phenomenology is not clear. But this form of writing the bottom-up
holographic models is known to be useful in discussions of confining aspects~\cite{br3}. We recall that the gravitational potential energy
for a body of mass $M$ is given by $V=M\sqrt{g_{00}}$ in units where $c=1$. The confinement behavior can be qualitatively deduced from following a particle in
warped AdS space as it goes to the infrared region, i.e. the region of large $z$, --- in general relativity, this would correspond to falling an object by the effects of gravity.
If the potential $V$ has an absolute minimum at some $z_0$ then one may speculate that a particle is confined in a hadron within distances $z\sim z_0$~\cite{br3}
(within the light-front approach, the holographic coordinate $z$ is proportional to the interquark distance in a hadron~\cite{br3}). This heuristic argument can be
substantiated by the Sonnenschein condition for Wilson loop-area law for confinement of strings: A background dual to a confining theory should satisfy
the condition~\cite{sonnenschein}
\begin{equation}
\label{sonn}
\partial_z(g_{00})|_{z=z_0}=0,\qquad g_{00}|_{z=z_0}\neq 0,
\end{equation}
which is nothing but the condition for extremum of the potential $V$.

Using Eq.~\eqref{warped} the time-time component of the warped metric of the SW model with positive and negative dilaton background, SW$^\pm$,
can be written as (we set $R^2|c|=1$ in what follows)
\begin{equation}
\label{w2}
\tilde{g}_{00}=\frac{e^{\pm|c|z^2(3/2-J)^{-1}}}{|c|z^2}.
\end{equation}
In the vector case $J=1$, the SW$^+$ model looks confining. The given observation underlies a traditional argument that the SW$^-$ model lacks confinement (see~\cite{br3} for a review).
This argument was substantiated by an explicit holographic derivation of linear confining potential via the Wilson loop in Ref.~\cite{AZ} using the warped metrics with positive exponent.
Actually the first SW model was introduced in Ref.~\cite{andreev} exactly in this form. We note, however, that the given conclusion formally changes to the opposite one in the
tensor case, for spins $J>1$ (the exponent in~\eqref{w2} changes the sign).

The range of possibilities in confining behavior becomes much wider in the case of generalized SW models introduced in the previous section.
The component $g_{00}$ in~\eqref{w2} is then generalized to
\begin{equation}
\label{w3}
\tilde{g}_{00}=\frac{\left[e^{-|c|z^2}U^2(b,J-1,|c|z^2)\right]^{(3/2-J)^{-1}}}{|c|z^2},\qquad J=0,1,
\end{equation}
\begin{equation}
\label{w4}
\tilde{g}_{00}=\frac{\left[e^{|c|z^2}M^2(-b,J-1,-|c|z^2)\right]^{(3/2-J)^{-1}}}{|c|z^2},\qquad J>1.
\end{equation}
The behavior of $\tilde{g}_{00}$ at different $J$ and $b\neq0$ can be analyzed graphically on a computer.
Our analysis showed that there are only two situations when a minimum appears, both happen for $b<0$.

\begin{figure}[!ht]
  \begin{minipage}[ht]{0.46\linewidth}
    \includegraphics[width=1\linewidth]{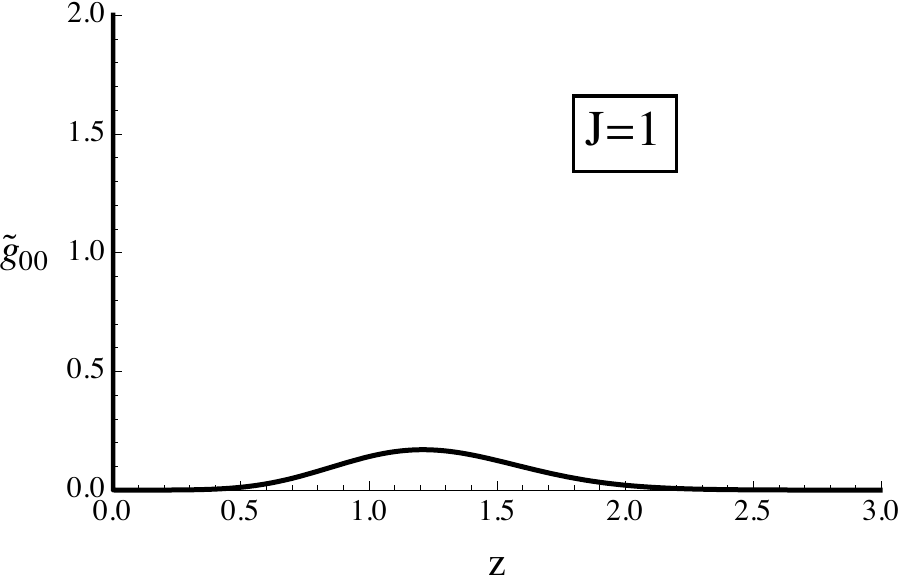} \\
  \end{minipage}
  \hfill
  \begin{minipage}[ht]{0.46\linewidth}
    \includegraphics[width=1\linewidth]{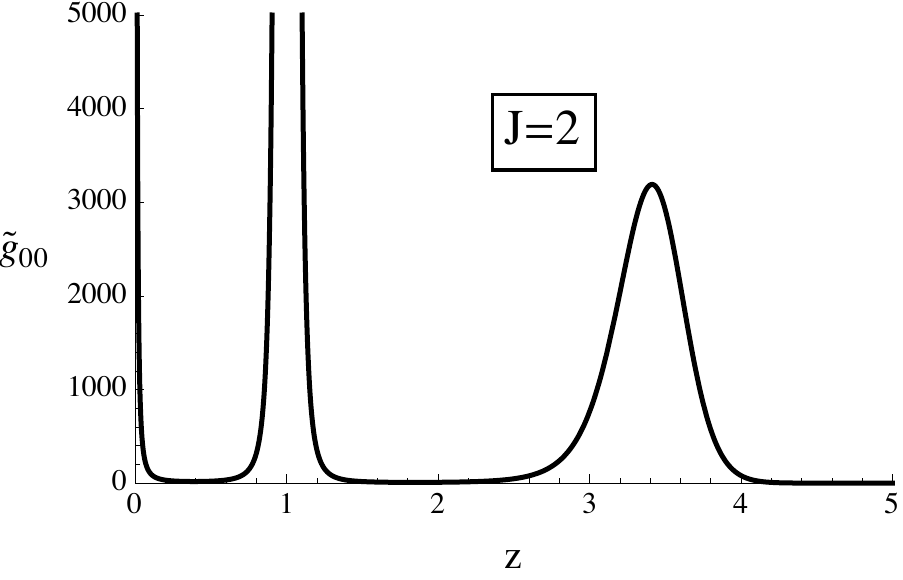} \\
  \end{minipage}
  \vspace{-0.8cm}
  \caption{\small The time-time component of metrics with \(|c|=1\): \eqref{w3} with
  \(J=1\) and \(b=-0.99\); \eqref{w4} with \(J=2\) and \(b=-1.99\).}
  \label{g00_plot}
\end{figure}

The first situation takes place for the vector $J=1$ case, where a local minimum emerges in some range of intercept parameter $-1<b<b_0$.
If we set $|c|=1$ then $b_0\approx-0.95$.
When $b\rightarrow-1$, the local minimum tends asymptotically to a kind of absolute
minimum situated at $z_0\rightarrow0$ and separated from the global minimum at $z\rightarrow \infty$ by a finite barrier, see Fig.~\ref{g00_plot}.
In the exact limit $b=-1$, however, the second condition in~\eqref{sonn} is not fulfilled.
On the other hand, the action~\eqref{genJ} cannot be extended to $b=-1$ since one has no AdS$_5$ space in the UV asymptotics
(as a consequence of $U(b,0,0)=1/\Gamma(1+b)$), hence, the holographic approach is not justified.

The second situation arises in the tensor $J=2$ case, where there is a local minimum for $-2<b<-1$. But in this case it is
separated from the global minimum at $z\rightarrow \infty$ by an infinite barrier at some $z_0$. This barrier emerges due to a specific singularity:
The Kummer function in~\eqref{w4} stays in negative power but this function has a zero at $z_0$, $M(-b,1,-|c|z_0^2)=0$. The root $z_0$ tends to
$1/\sqrt{|c|}$ when $b$ tends to $-2$, see Fig.~\ref{g00_plot}.

The limit $b\rightarrow-1$ in the \(J=1\) case corresponds to the limit of massless vector particle in the spectrum~\eqref{primary2}.
As is clear from the discussion above, we can find this particle only by taking the limit $b\rightarrow-1$, not just setting $b=-1$
(this somewhat resembles the prescription $\varepsilon\rightarrow0$ in calculating QFT Green functions)
and only in a deep UV region $z_0\rightarrow0$. The presented construction could thus serve as a basis for a holographic
realization of asymptotic freedom of gluons in QCD.

In the limit of \(b\to-2\) of the $J=2$ case, another singularity gradually appears. In addition, this limit corresponds to the limit of massless spin-2 particle in the spectrum~\eqref{primary2}.
For this reason we do not discuss the minima appearing at $b<-2$, they would be related to unphysical tachyonic states in the spectrum~\eqref{primary2}.
In the parametric interval $-2<b<-1$, we get a confinement of spin-2 particles with respect to propagation to deep infrared domain.

\subsection{Chiral symmetry breaking}

The strong interactions reveal two separated mass scales, $\Lambda_\text{QCD}\approx0.2$~GeV and $\Lambda_\text{CSB}\approx1$~GeV.
The scale $\Lambda_\text{QCD}$ is related to confinement while $\Lambda_\text{CSB}$ characterizes the spontaneous Chiral Symmetry Breaking (CSB).
The linear Regge spectrum also depends on two parameters, the slope and intercept. Experimentally, the meson slope is almost universal in the
light quark sector~\cite{ani,bugg,klempt,phen,phenSV,phen2,phen3,phen4,phen5} while the intercept depends on the spatial parity of radial meson
trajectory~\cite{phen2,phen3,phen4,phen5}, a short review is given in the Appendix~D. It is thus natural to assume that the slope
is proportional to $\Lambda^2_\text{QCD}$ while the intercept is somehow connected to $\Lambda_\text{CSB}$.

A satisfactory description of the CSB within the SW approach and its extensions is an open problem.
One can approach this problem by gradually reproducing various effects and features of CSB.
The first effect is a generation of almost massless pion. The second effect consists in a large mass splitting
between chiral partners. The given two effects must be somehow related. This can be easily tuned in the generalized SW
model under consideration. Indeed, consider, e.g., the mass spectrum~\eqref{vect3} and take $c>0$.
Since both the vector and scalar quark operators in QCD have minimal dimension $\Delta=3$, the relation~\eqref{3c} yields
$m_5^2=0$ for vectors and $m_5^2R^2=-3$ for scalars. The spectrum~\eqref{vect3} then takes the form
\begin{equation}
\label{J1}
J=1:\qquad m_n^2=4c(n+1+b),
\end{equation}
\begin{equation}
\label{J0}
J=0:\qquad m_n^2=4c(n+1/2+b).
\end{equation}
One can further assume that due to some mechanism the intercept parameter $b$ is different for mesons
possessing negative and positive parity, $b_-\neq b_+$, and that $b_-\approx-1/2$. We then get
approximately massless pion and simultaneously obtain the radial spectrum~\eqref{12lin} for the $\rho$-mesons
which is often encountered in the literature, see the comments below.
If $b_+\approx0$ we arrive at the well-known pattern of CSB in the meson spectrum (see, e.g., discussions in Refs.~\cite{gutsche,lin}).
In particular, for isospin-one light non-strange vector and axial-vector mesons this pattern yields the following radial spectrum,
\begin{equation}
\label{12lin}
m_\rho^2(n)=2m_\rho^2\left(n+\frac12\right),
\qquad
m_{a_1}^2(n)=2m_\rho^2\left(n+1\right),
\end{equation}
where $m_\rho$ is the mass of the ground $\rho$-meson. For the ground states, $n=0$, the mass splitting
between the opposite parity vector mesons can be matched to a prediction of low-energy quark models
based on the effective four-fermion interactions~\cite{klev} (in the framework of which the concept of spontaneous CSB first emerged)
\begin{equation}
\label{ma}
m_\rho^2=m_{a_1}^2-6M_q^2= 2m_\rho^2\left(1-\frac{6M_q^2}{2m_\rho^2}\right),
\end{equation}
where
\begin{equation}
\label{cqm}
M_q\sim-\langle\bar{q}q\rangle,
\end{equation}
is the constituent quark mass, $M_q\approx320$~MeV. Strictly speaking, $M_q\sim-\langle\bar{q}q\rangle/\Lambda^2$, where
$\Lambda$ represents the UV cutoff in a model.
If $b_+\approx0$ then we get
\begin{equation}
\label{bminus}
b_-\approx-\frac{3M_q^2}{m_\rho^2}\approx-\frac12.
\end{equation}
The given numerical agreement substantiates the assumption that the intercept $b$ arises from the spontaneous CSB.

Digressing for a moment, we should mention that
the relations~\eqref{12lin} are very old and were emerging in various approaches in the past.
First they appeared in the extensions of Veneziano dual amplitude for $\pi+\pi\rightarrow\pi+\pi$ scattering
to the reactions $\pi+A\rightarrow B+C$, the obtained positions of poles were~\cite{avw,avw2,collins}
\begin{equation}
\label{ls1}
\text{$\rho$-trajectory}:\qquad m^2_{(\rho)}=2m_{\rho}^2(n+J-1/2), \qquad J=1,2,\dots,
\end{equation}
\begin{equation}
\label{ls2}
\text{$\pi$-trajectory}:\qquad m^2_{(\pi)}=2m_{\rho}^2(n+J), \qquad J=0,1,\dots.
\end{equation}
where $n=0,1,2\dots$ enumerates the daughter Regge trajectories (see the Appendix~E). For $J=0$ in the spectrum~\eqref{ls1}, one has to substitute $J=1$,
i.e. the scalar and vector mesons turn out to be degenerate in mass (the given prediction is not unreasonable, see Fig.~\ref{OmegaF} in the Appendix~E).
The shift of Regge trajectories corresponding to opposite parities
arises from the Adler self-consistency condition~\cite{avw,avw2,collins} imposed on the Veneziano amplitude ---
the amplitude of $\pi\pi$ scattering must be zero at zero momentum transfer. This condition
incorporates the CSB at the amplitude level and automatically removes degeneracy between the $\rho$ and $a_1$ spectra.
The corresponding ideas are briefly reviewed in the Appendix~E.

Also the relations~\eqref{12lin} were derived in the framework of finite energy QCD sum rules in Ref.~\cite{Krasnikov}.
Later it was shown that the vector spectrum~\eqref{12lin} represents a general result of QCD sum rules when the large-$N_c$ limit is
taken and linearity is imposed~\cite{lin}. The values of intercept $b=0$, $b=1/2$ and $b=1$ turned out to be distinguished by the property that
a half of condensates disappears in the OPE of two-point vector correlators. Various aspects of this result were analyzed in Ref.~\cite{lin2,lin2b}.
Remarkably, the $\rho$-spectrum in~\eqref{12lin} realizes a minimum of non-perturbative contributions from condensates to the parton model logarithm in the OPE,
in this sense, it is maximally dual to the perturbative continuum\footnote{If the Regge behavior (linearity of masses squared) is not imposed, the parton model logarithm is
approximated by an infinite sum of poles in the "best possible way" when these poles are given by zeros of Bessel function $J_0$~\cite{migdal}. It looks astonishing
that this spectrum obtained by Migdal in 1978 from a kind of quark-hadron duality requirement
was reproduced by the first Hard Wall holographic models~\cite{son1,pom}, see discussions in Ref.~\cite{low}.}~\cite{lin3}.
Recently the spectral relations~\eqref{12lin} were reported in the extended quark model of Ref.~\cite{ahep2}.
In a particular case of $n=0$, $J=1$, the spectrum~\eqref{12lin} leads to the famous Weinberg relation~\cite{wein},
$m_{a_1}^2=2m_{\rho}^2$. The given relation expresses the fact of maximal mixing of
longitudinal component of axial $a_1$-meson with the pion after CSB,
while between the vector and scalar states such a mixing is absent~\cite{gh,gh2}.

Returning to our subject matter, the next step should be an explicit incorporation of the aforementioned CSB effect into the holographic SW approach
in a form of some dynamical mechanism.
An interesting variant for such a mechanism was suggested within the Light-Front (LF) holographic QCD~\cite{br3}.
The LF holographic approach prescribes $\Delta= L+2$ for the twist-two QCD operators, where $L$ represents
the maximal value of the $z$ component of the quark orbital angular momentum in the LF wave function.
Substitution of this prescription into the relation~\eqref{3c} yields
\begin{equation}
\label{LFm}
\text{LF}: \qquad m_5^2R^2=L^2-(J-2)^2.
\end{equation}
This prescribes $\sqrt{(J-2)^2+m_5^2R^2}=L$ in the holographic bound state equation~\eqref{primary}
leading finally to the spectrum
\begin{equation}
\label{LF}
\text{LF}:\qquad m_n^2=4c\left(n+\frac{J+L}{2}\right).
\end{equation}
The spectrum~\eqref{12lin} follows immediately as a particular case of Eq.~\eqref{LF}~\cite{br3,gutsche}.

In reality, however, the experimental spectrum of light mesons is not well fitted neither by the standard SW relation~\eqref{string},
even in our extended variant,
\begin{equation}
\label{ext}
\text{Extended SW}:\qquad m_n^2=4c\left(n+J+b\right),
\end{equation}
nor by the LF one\footnote{When $L$ is interpreted as the orbital momentum of an quark-antiquark system following the
standard assignments of Particle Data~\cite{pdg}, as in the fit~\eqref{hydr}. We carried out a phenomenological analysis with the aim of testing the relation~\eqref{LF},
the results of our global fit were unsatisfactory.}~\eqref{LF}. A good fit of data within the linear ansatz for masses square
is given by the relation~\cite{phen4,phen5}
\begin{equation}
\label{hydr}
\text{Experiment}:\qquad m_n^2=a\left(n+L+b\right), \quad a\approx1.1\,\text{GeV}^2, \quad b\approx0.6,
\end{equation}
see the Appendix~D for a more detailed discussion.
It looks like in Eq.~\eqref{ext} we should take the intercept parameter
\begin{equation}
\label{b2}
b=L-J+1/2.
\end{equation}
We do not know how to derive this prescription, but at least a way for further development can be delineated clearly:
A SW like holographic model correctly describing the global CSB effect in the meson spectrum should lead to an intercept close to~\eqref{b2}.
For instance, for vector and axial mesons the prescription~\eqref{b2} gives $b_\pm=\pm1/2$. It is interesting to note that the generalized SW
model with non-zero intercept after matching to the Operator Product Expansion of two-point correlators results in the pattern $|b_+|=|b_-|$~\cite{genSW}
(but quantitatively one obtains $|b_\pm|\approx1/3$ from the phenomenological value of gluon condensate~\cite{genSW}).

Consider a concrete phenomenological example ---
the non-strange iso\-singlet states in the vector and axial-vector channels, the $\omega$ and $f_1$ mesons
(the experimental spectrum of $\rho$-meson resonances contains some contradictory superfluous states~\cite{pdg},
we wish to avoid complications).
The Particle Data~\cite{pdg} reports three well-established
$\omega$-mesons: $\omega(782)$, $\omega(1420)$, and $\omega(1650)$.
Taking their masses from~\cite{pdg} and ascribing them the radial
numbers $n=0,1,2$, we get the fit (in GeV$^2$): $m^2_{\omega}(n)\approx1.1(n+0.7)$.
The axial sector contains only one well-established
state $f_1(1285)$ (another one, $f_1(1420)$, consists mostly of the
strange quarks). Let us use the non-confirmed
states $f_1(1970)$ and $f_1(2230)$~\cite{pdg} as a guess. We ascribe them
the radial numbers $n=2,3$ (a state corresponding to
$n=1$ --- the isoscalar partner of $a_1(1640)$~\cite{pdg} --- is not
established). This results in the fit: $m^2_{f_1}(n)\approx1.1(n+1.5)$.
The slopes in the obtained linear fits agree with~\eqref{hydr}, the magnitudes of intercepts
are close to the aforementioned pattern $b_\pm=\pm1/2$.

The CSB is probably responsible for a specific phenomenon which is not reflected in the mass formula~\eqref{hydr}:
The ground states in the vector and pseudoscalar channels lie significantly below the corresponding linear radial Regge trajectories.
It means that if a radial trajectory is fitted using only the excited states and after that the mass of ground state is predicted,
the prediction will appreciably overestimate the real mass (see, e.g., the discussions in Ref.~\cite{srAS} and the
Fig.~\ref{OmegaF} in the Appendix~E). It is interesting to remark that in the vector mesons, where there are many data
both in the light and heavy quark sectors, this effect is almost independent of quark mass~\cite{phen6,phen6b}.
As a result, the spectrum~\eqref{12lin} does not give a good fit for the radially excited $\rho$-mesons ---
the ground state must be separated and the intercept should be taken noticeably larger.
This is visualized in Fig.~\ref{OmegaF} of the Appendix~E for the case of $\omega$-mesons.
The same can be repeated for pseudogoldstone bosons. For instance, the spectrum
\begin{equation}
\label{b12lin}
m_\pi^2(n)=2m_\rho^2 n,
\end{equation}
which would naively follow from the Eq.~\eqref{LF}, does not fit well the masses of radially excited pions.

The ground state $n=J=L=0$ in the Eq.~\eqref{LF}, which is associated with massless pion in the LF holographic QCD, deserves a special comment:
The Schr\"{o}dinger equation~\eqref{sr3} (where $s=L$ in the given case) does not have a discrete spectrum for $L=0$ as the potential
is not bounded from below. The spectrum~\eqref{LF} therefore cannot be extrapolated to the massless case.

As far as we can see, the problem is rooted in the change of constant part of 5D mass when the prescription~\eqref{LFm} is imposed.
This would physically mean that the CSB is a local effect occurring in the deep UV domain. But the effect of spontaneous CSB is known
to be essentially non-local. The non-local effects are naturally incorporated in our generalized SW model via the effective $z$-dependent
mass~\eqref{meff}, let us write once more its structure,
\begin{equation}
\label{meff2}
m_5^2(z)=A+Bz^2+Cz^4.
\end{equation}
The constant part $A$ is given by the AdS/CFT prescription~\eqref{3c} in the deep UV domain $z\rightarrow0$ and should not be changed.
At some non-zero distances, certain non-local effects appear which are modeled by the $\mathcal{O}(z^2)$ term and supposedly related to the spontaneous CSB.
At even larger distances, the $\mathcal{O}(z^4)$ term comes into play and describes confinement (leading to universal linear Regge trajectories).
The intercept parameter $b$ in~\eqref{J1} and~\eqref{J0} originates both from $A$ and $B$ in the effective mass~\eqref{meff2}. An advantage of the proposed generalized
SW approach is that we can keep $A$ constant and vary $B$. We believe that this gives a physically more plausible direction for description of CSB due to
incorporated non-locality and, as a consequence, absence of serious problems with having a massless scalar state.

The traditional way for phenomenological description of the CSB in the bottom-up holographic approach is based on
condensation of a scalar field which dynamically acquires a $z$-dependent ``vacuum expectation value'' in a form of non-normalizable
and momentum-independent solution to the scalar equation of motion~\cite{son1,pom}. The pseudogoldstone mesons are introduced, in essence,
in the same way as in the chiral perturbation theory and related effective low-energy models\footnote{Within the Hard Wall holographic models,
there exists an alternative approach, in which the mass splitting between vector and axial mesons arises from different boundary conditions
imposed on the corresponding 5D fields and the pion represents the fifth component of 5D axial field~\cite{Hirn:2005nr}. Phenomenologically the given
approach was less successful than the traditional one.}. It is known, however, that this scheme
cannot be directly realized in simple SW models~\cite{son2}. The underlying reason is that one of two solutions to the equation of motion
has incompatible divergent behavior and should be discarded\footnote{It is interesting to mention a recent solution proposed in Ref.~\cite{EKpion}
which was based on embedding $\mathcal{O}(z^2)$ corrections to the 5D masses, the given proposal is in line with our more general approach.}.
But this leads to a wrong pattern of CSB (the quark condensate is proportional to the current quark mass\footnote{An alternative solution can consist in
the assumption (see Ref.~\cite{LowAdS}) that if an IR modification is incorporated into an AdS/QCD model then the corresponding source term,
the first term in the Eq.~\eqref{25a}, may be interpreted as the constituent quark mass which is proportional to the the quark condensate,
see the relation~\eqref{cqm}.}). A more realistic description of CSB in the SW models is usually achieved at the cost of various
extensions with new free parameters.

In the present holographic setup, the standard description of the CSB may be outlined as follows.
As we have demonstrated in our analysis above,
the most general and compact (i.e., not including any additional fields) SW model leading to exactly linear spectrum is defined by
the $z$-dependent effective mass~\eqref{meff2}. All other formulations can be regarded as its derivatives. If one introduces a 5D scalar
field $X$ dual to the quark condensate, nothing precludes to set $C=0$ and (but not necessary) $B=0$ in~\eqref{meff2}.
One may try then to construct the description of CSB in the same way as in the Hard Wall holographic model~\cite{son1,pom}
(the GOR relation for pion mass, mass splittings, various decay constants and constants of chiral perturbation theory).
In terms of the exponential dilaton background, this would mean the following: A SW model incorporating the field $X$ should have the form
\begin{equation}
\label{31c}
S_X=\int d^4\!x\,dz\sqrt{g}\left(e^{cz^2}\!\mathcal{L}+\partial^M\! X\partial_M\! X-m_X^2 X^2\right),
\end{equation}
where the Lagrangian density $\mathcal{L}$ contains all fields under study and their interactions with $X$.
The effect of confinement described by the background $e^{cz^2}$ and the effect causing the CSB (occurring at substantially
smaller distances) become clearly separated in the first approximation.
It is worth mentioning in passing that a similar to~\eqref{31c}
construction with $e^{cz^2}$ replaced by $e^\varphi$ and $X$ by $\varphi$,
\begin{equation}
\label{31d}
S_\varphi=\int d^4\!x\,dz\sqrt{g}\left(e^{\varphi}\!\mathcal{L}+\partial^M\! \varphi\partial_M\! \varphi-m_\varphi^2 \varphi^2\right),
\end{equation}
was often used in the literature to motivate the origin of dilaton background: If $\Delta=2$, i.e. $m_{\varphi}^2R^2=-4$ in
the equation~\eqref{24a} for dilaton field $\varphi$, this equation has a solution $\varphi\sim z^2$.

Following the original papers~\cite{son1,pom}, one should consider
the equation of motion for $X$ in the absence of other fields and independent of usual 4D coordinates (equivalently, 4D momentum $q$),
\begin{equation}
\label{24a}
-\partial_z\left(\frac{\partial_z X}{z^3}\right)+\frac{m_{X}^2R^2X}{z^5}=0.
\end{equation}
The scalar quark bilinear operator $\bar{q}q$ in QCD has canonical dimension
$\Delta=3$, the UV prescription~\eqref{3c} dictates $m_{X}^2R^2=-3$ that results in the solution
\begin{equation}
\label{24b}
X=c_1z+c_2z^3.
\end{equation}
According to general AdS/CFT rules, the near-boundary solution of the equation of motion
for a field $\Phi^\mathcal{O}$ dual to an operator $\mathcal{O}$ with dimension $\Delta$
is given by~\cite{kleb}
\begin{equation}
\label{25a}
\Phi^\mathcal{O}(z,q)_{z\rightarrow0}=z^{4-\Delta-J}\Phi_0^\mathcal{O}(q)+z^{\Delta-J}\frac{\langle \mathcal{O}\rangle}{2\Delta-4},
\end{equation}
where $\Phi_0^\mathcal{O}(q)$ represents the source for $\mathcal{O}$ and $\langle\mathcal{O}\rangle$ is the corresponding condensate
(the one-point correlation function).
The comparison of prescription~\eqref{25a} for $\Delta=3$ and $J=0$ with the solution~\eqref{24b} leads, as usual, to
interpretation of $c_1$ as a constant proportional to the current quark mass $m_q$ and $c_2$ becomes proportional to
the quark condensate $\langle\bar{q}q\rangle$.

As was demonstrated in Refs.~\cite{kwee1,kwee2}, the further phenomenology can be developed in line with the pioneering
papers~\cite{son1,pom} if the SW dilaton background is negative, $c<0$. Actually, the form of solution~\eqref{24b} was taken
in~\cite{kwee1} by hand and in~\cite{kwee2} it was shown that numerical deviations from exact form in typical extended versions of SW model
(with $c_1$ and $c_2$ being independent) are small due to arising exponential damping. Thus a somewhat speculative analysis of Ref.~\cite{kwee1}
becomes completely justified if the ansatz~\eqref{31c} is accepted. The cases of $c>0$ and $c=0$ (the No-wall model with a $z$-dependent 5D mass)
are problematic since the Langrangian for normalizable mode of $X$ diverges. It seems that some extra assumptions or modifications are needed
in order to advance in these cases. We left this as a possible work for the future.

It should be emphasized that the outlined standard way for the description of CSB is very limited in scope:
It can describe the CSB in the vector and axial channels but the phenomenology shows that the same phenomenon of CSB recurs for higher spin
mesons. In the Appendix~D, we recall the known observation that not only the $\rho$ and $\omega$ mesons but also all mesons belonging to the
leading $\rho$ and $\omega$ Regge trajectories do not have parity partners. {\it The spontaneous CSB affects the whole Regge trajectory!}
For instance, it leads to a splitting
$m_{a_3}^2-m_{\rho_3}^2$ which is approximately the same as the splitting $m_{a_1}^2-m_{\rho}^2$ described in the standard CSB scenario.
Note in passing that the LF holographic QCD describes this important feature automatically, see the spectrum~\eqref{LF}.
An extension of the standard scenario to the tensor mesons, however, becomes problematic because this would involve local interactions of
the field $X$ with the higher spin fields. But this would lead to a notorious problem with violation of causality,
a universally applicable self-consistent theory for locally interacting tensor fields is not known.

Within our phenomenological
approach, we can avoid this problem by simply postulating an effective infrared $\mathcal{O}(bz^2)$ contribution to the 5D mass.
The most economical possibility is to take a universal intercept parameter for all mesons with equal parity, i.e. to introduce two new parameters,
$b_-$ and $b_+$ (one of them can be zero). We get then a universal shift of the whole Regge trajectory caused by the CSB. Exactly in this way
the CSB affects the spectrum in the Veneziano like dual amplitudes, see the Appendix~E.

On the phenomenological level, the $\mathcal{O}(z^2)$ contribution to the 5D mass can be introduced via the interaction term
\begin{equation}
\label{intX}
\mathcal{L}_\text{int}\sim bX^2_\text{non}\Phi^{M_1\dots}\Phi_{M_1\dots},
\end{equation}
where $X_\text{non}$ denotes the non-normalizable $\mathcal{O}(z)$ solution. Since non-normali\-zab\-le
solutions in the holographic approach correspond to some background, one can speculate that the term~\eqref{intX} describes
a non-local interaction of tensor fields with a background responsible for the CSB. Choosing a proper normalization of 5D fields,
the intercept parameter $b$ becomes a coupling. A simple requirement that this coupling is different for different parities would
lead to the scenario discussed above. It should be added finally that the $\mathcal{O}(z^4)$ contribution to the 5D mass can be
introduced along the same line --- in this case, $X_\text{non}$ should correspond to some dimension-two condensate as in the scenario
of Ref.~\cite{No-wall}.

\subsection{Two-point vector correlator}

The central objects in the holographic approaches based on the AdS/CFT correspondence are correlation functions of various operators.
The famous AdS/CFT prescription of Refs.~\cite{witten,gub} provides a recipe of calculating these functions from a dual gravitational theory.
Perhaps the most remarkable aspect is that when the dual higher dimensional theory can be treated semiclassically, i.e. as just a classical
field theory in a curved space, the obtained correlation functions correspond to the strong coupling regime of 4D gauge theory due to
the strong-weak nature of duality in the gauge/gravity correspondence.

The two-point correlation function of vector currents $J_{\mu}$ is defined in the Euclidean space as
\begin{equation}
\label{cor13}
\int d^4x e^{iqx}\langle
J_{\mu}(x)J_{\nu}(0)\rangle=(q_{\mu}q_{\nu}-q^2g_{\mu\nu})\Pi_V(Q^2), \qquad Q^2=-q^2.
\end{equation}
Applying the standard AdS/CFT prescription~\cite{witten,gub} to the bottom-up holographic models, the vector
two-point correlator $\Pi_V(Q^2)$ is determined by the expression~\cite{son1,pom} (we set to unity the general normalization constant)
\begin{equation}
\label{vc2}
  \Pi_V(Q^2) = \left.-\frac{\partial_z V(Q^2,z)}{Q^2z}\right|_{z\rightarrow0}.
\end{equation}
where $V(Q^2,z)$ represents the so-called bulk-to-boundary propagator.
Such propagators are of primary importance in the holographic approaches because they determine the correlation functions.
In the vector case under consideration, the function $V(Q^2,z)$ is defined
as the solution of equation of motion for spin-one mesons in the Euclidean domain
satisfying the boundary condition
\begin{equation}
V(Q^2,0)=1.
\end{equation}
Within the standard SW holographic model with the background $e^{cz^2}$ , the corresponding solution reads~\cite{son3}
\begin{equation}
\label{btb}
  V(Q^2,z)=\Gamma\lb 1+ \tilde{Q}^2\rb e^{-(c+|c|)z^2/2} U\lb  \tilde{Q}^2,0,|c|z^2\rb,
\end{equation}
where the dimensionless momentum,
\begin{equation}
\label{tildeQ}
\tilde{Q}^2\equiv\frac{Q^2}{4|c|},
\end{equation}
is introduced for simplicity.
Actually after extracting the prefactor $e^{-cz^2/2}\sqrt{z}$ the finding of this solution just repeats the finding
of generalized background function $f_1(z)$ in~\eqref{sol1} with $b$ replaced by $\tilde{Q}^2$, i.e., $V(Q^2,z)$
can be written from~\eqref{sol1} by setting $J=1$, $b=\tilde{Q}^2$ and multiplying by that prefactor.
The expansion of~\eqref{btb} near $z=0$ reads
\begin{multline}
\label{expan}
  V(Q^2,z)_{z\rightarrow 0}=1+\\
\left\{\tilde{Q}^2\left[\ln(|c|z^2)
+\psi\left(1+\tilde{Q}^2\right)+2\gamma-1\right]-\frac12\left(1+\frac{c}{|c|}\right)\right\}|c|z^2.
\end{multline}
The vector two-point correlator of the SW model (first derived for general $c$ in~\cite{son3}) follows
from substitution~\eqref{expan} into~\eqref{vc2},
\begin{equation}
\label{vc}
  2\Pi_V(Q^2) = \frac{1+\frac{c}{|c|}}{2\tilde{Q}^2}-\psi\left(1+\tilde{Q}^2\right)+\text{Const},
\end{equation}
where $\psi$ denotes the digamma function that can be represented as a sum of pole terms,
\begin{equation}
\label{11a}
\psi\left(1+\tilde{Q}^2\right)=-\sum_{n=0}^{\infty}\frac{1}{\tilde{Q}^2+n+1}+\text{const}.
\end{equation}
The poles
\begin{equation}
\label{polM}
-\tilde{Q}^2=\frac{m_n^2}{4|c|}=n+1, \qquad n=0,1,\dots,
\end{equation}
of this function yield the mass spectrum that can be also found by solving the corresponding equation of motion.
A convenient visualization of the origin of pole structure is given by the following representation of
bulk-to-boundary propagator~\eqref{btb} derived in Ref.~\cite{Radyushkin2007},
\begin{equation}
\label{repr}
V(Q^2,z)=4c^2z^2e^{-(c+|c|)z^2/2}\sum_{n=0}^{\infty} \frac{L_n^1(|c|z^2)}{Q^2+4|c|(n+1)}.
\end{equation}
The poles and residues in~\eqref{repr} yield directly the spectrum~\eqref{primary} and eigenfunctions~\eqref{wf2} for $J=|s|=1$.

As was emphasized in Ref.~\cite{son3}, the choice of positive
dilaton background, $c>0$, leads to unphysical massless pole in the vector correlator~\eqref{vc}, the physical choice is the sign $c<0$.
This conclusion is in accord with our result found by shifting the intercept parameter $b$ from
zero value (assumed by default in the standard SW models) with the help of modification of SW background:
The generalized background in SW action~\eqref{genJ} reducing to the SW$^-$ model
at $b=0$ is well defined for the scalar and vector case only.

On the other hand, we can redefine the SW model introducing the $\mathcal{O}(c^2z^4)$ infrared modification of 5D mass~\eqref{meff2}
instead of the exponential background $e^{cz^2}$. This is equivalent to setting $c=0$ but keeping $|c|\neq 0$ in all relevant relations
including~\eqref{btb} as was first shown in Ref.~\cite{No-wall}. This is qualitatively clear --- the results cannot depend on the sign of $c$
within such a ``No-wall'' formulation of the SW model. Below the given formulation will be referred to as SW$^0$ model.
The definitions can be briefly summarized as
\begin{equation}
\text{SW}^\pm: \qquad \text{Background}\,\,e^{\pm|c|z^2},\quad m_5^2R^2=0.
\end{equation}
\begin{equation}
\text{SW}^0: \qquad \text{Background}\,\,e^{0},\quad m_5^2R^2=c^2z^4.
\end{equation}
It is seen that the massless pole remains in the SW$^0$ model and already cannot be removed.
In addition, if we introduce arbitrary intercept parameter $b$ with the help of $\mathcal{O}(bz^2)$ infrared modification of 5D mass~\eqref{meff2},
the massless pole emerges even for the negative sign $c<0$~\cite{AS}, see below. All this looks troublesome as long as the proportionality
of two-point function~\eqref{cor13} to $q^2=-Q^2$ is necessary for gauge invariance. The crux of the problem is that the $Q^2\Pi_V(Q^2)$
defines the hadronic vacuum polarization and would enter a dressed photon propagator, making photons massive if it does not vanish at $Q^2=0$.
In other words, we must have the condition
\begin{equation}
\label{condit}
 \left. Q^2\Pi_V(Q^2)\right|_{Q^2=0} = 0.
\end{equation}

The condition~\eqref{condit} is quite non-trivial in the general case.
The direct substitution of bulk-to-boundary propagator~\eqref{btb} to~\eqref{vc2} leads to a UV divergent expression.
The divergent constant is usually ignored together with a finite constant in~\eqref{vc} because only the pole structure is typically
of interest. In the simple SW$^-$ model, this is justified by a rigorous holographic renormalization~\cite{skenderis}.
We recall that the correlator $\Pi_V(Q^2)$ is defined in such a way that the transverse projector in the polarization operator~\eqref{cor13} is proportional to $Q^2=-q^2$.
This definition is usually accepted keeping in mind the property~\eqref{condit} due to the gauge invariance. 
But following the holographic prescriptions one finds originally an expression for the object ``$Q^2\Pi_V(Q^2)$''.
Let us ``restore'' the general factor $Q^2$ and write the explicit answer for the renormalized two-point correlator in the usual SW model,
\begin{equation}
\label{vc3}
  Q^2\Pi_V(Q^2) = |c|\left\{1+\frac{c}{|c|}-2\tilde{Q}^2\left[\psi\left(1+\tilde{Q}^2\right)+2\gamma-1+C_\text{ct}\right]\right\},
\end{equation}
where the constant $C_\text{ct}$ stems from the local counterterm required for cancelation of UV divergence. 
Since this term is proportional to the square of the gauge invariant field strength, it can appear only with a factor $Q^2$.
The constant $2\gamma-1$ appears from the expansion~\eqref{expan}. This constant plus $C_\text{ct}$ give the constant contribution
``Const'' in the expression~\eqref{vc}. It is seen that the counterterm contribution $C_\text{ct}$ cannot cancel
the constant $1+\frac{c}{|c|}$. The latter constant is zero only in the SW$^-$ model, 
the renormalization of simple SW$^+$ and SW$^0$ models cannot remove the unphysical massless pole. 

The situation becomes more interesting if the expression~\eqref{vc3} is extended by inclusion of the intercept parameter $b$. 
Following our results summarized in Section~4.3, the arbitrary intercept can be introduced into any variant of SW model
by adding $\mathcal{O}(bz^2)$ infrared modification of 5D mass~\eqref{meff2},
this is equivalent just to the shift $\tilde{Q}^2\rightarrow\tilde{Q}^2+b$ in the corresponding equation of motion
(in our normalization of intercept, i.e. as in the Eq.~\eqref{vect3}).
Neglecting the counterterm, the direct extension of~\eqref{vc3} reads then as follows
\begin{equation}
\label{vc4}
  Q^2\Pi_V(Q^2) = |c|\left\{1+\frac{c}{|c|}-2(\tilde{Q}^2+b)\left[\psi\left(1+\tilde{Q}^2+b\right)+2\gamma-1\right]\right\}.
\end{equation}
The poles of this correlator are situated at
\begin{equation}
\label{polM2}
-\tilde{Q}^2=\frac{m_n^2}{4|c|}=n+1+b, \qquad n=0,1,\dots.
\end{equation}
Adding the counterterm contribution (which as usual is proportional to $Q^2$) the extension~\eqref{vc4} can be written as
\begin{multline}
Q^2\Pi_V(Q^2) = -2|c|\tilde{Q}^2\left[\psi\left(1+\tilde{Q}^2+b\right)+2\gamma-1+C_\text{ct}\right]\\
+|c|\left\{1+\frac{c}{|c|}-2b\left[\psi\left(1+\tilde{Q}^2+b\right)+2\gamma-1\right]\right\}.
\label{vc5}
\end{multline}
Now the condition~\eqref{condit} can be satisfied at certain values of intercept $b$, namely given by the algebraic equation
\begin{equation}
\label{eqb}
2b\left[\psi\left(1+b\right)+2\gamma-1\right]=1+\frac{c}{|c|}.
\end{equation}
The equation~\eqref{eqb} has the following numerical solutions satisfying $b\geq-1$
(this restriction excludes tachyonic states in the spectrum~\eqref{polM2}),
\begin{align}
\label{eqb1}
\text{SW}^-:& \qquad b_1^-=0;\quad b_2^-\approx0.31.\\
\label{eqb2}
\text{SW}^0:& \qquad b_1^0\approx-0.38;\quad b_2^0\approx0.94.\\
\label{eqb3}
\text{SW}^+:& \qquad b_1^+\approx-0.52;\quad b_2^+\approx1.31.
\end{align}

The solution $b_1^-$ corresponds to the known situation of absence of pole at zero momentum in the correlator~\eqref{vc}.
But as we see, a simple extension of the SW model to free intercept in the linear spectrum gives rise to five new possibilities.
It turns out that the SW$^+$ and SW$^0$ holographic models can be constructed for vector mesons without unphysical massless pole.
This solves, in particular, the longstanding problem of SW$^+$ model pointed out in Ref.~\cite{son3} (and noticed even earlier ---
in the pioneering paper~\cite{son2} where the SW$^-$ model was introduced).

A question appears about a possible physical meaning of the second solution in~\eqref{eqb1}~---~\eqref{eqb3}.
One can interpret the second solution as a prediction of the second radial trajectory with a larger intercept. Such an additional
trajectory for vector mesons is well known in the phenomenology --- it corresponds to the $D$-wave vector resonances~\cite{ani,bugg}.
In the light non-strange mesons, the ground states on these trajectories are $\rho(1700)$ for isotriplet and $\omega(1670)$
for isosinglet mesons~\cite{pdg}. Since the spectrum of light mesons reveals the global behavior~\eqref{hydr}, $m_n^2\sim n+L$,
the expected difference of intercept between the $S$-wave and $D$-wave trajectories is $\Delta b=2$. The prediction~\eqref{eqb3} of
SW$^+$ model, $\Delta b^+\approx1.83$, is close to this expectation. One can further suggest that the deviation of $\Delta b^+$
from $\Delta b$ is due to a lower value of slope of angular trajectories with respect to the slope of radial trajectories.
We get then a holographic prediction of this effect together with its magnitude. The latter is comparable with the fits of light meson
spectra performed in Ref.~\cite{arriola} (10\% in the SW$^+$ model vs. 20\% in~\cite{arriola}).

The solution $b_1^+$ is remarkably close to the
intercept~\eqref{b2} in the vector case (i.e., when $J=1$ and $L=0$) which is favored by the phenomenology and by other approaches.
And it turns out to be fully consistent with the fact that both~\eqref{b2} and the LF holographic spectrum~\eqref{LF} were derived
within the framework of SW$^+$ holographic model.

In the general situation of arbitrary intercept parameter, the holographic renormalization of extended SW model is problematic
because of appearance of longitudinal part in the two-point vector correlator as
in the SW$^+$ variant of the model at zero intercept.
Thus, the extended SW holographic model should be used in phenomenological applications keeping in mind this 
problem\footnote{We note, however, that in numerous phenomenological applications of SW$^+$ model, this problem is ignored.}.
On the other hand, the given problem may give impetus to further development. For instance, one can try to construct a physical mechanism
for cancelation of longitudinal contribution to the polarization operator. Or propose a clear physical interpretation for the longitudinal 
contribution (perhaps the most obvious option is that the photon excitations are prone to exist in a strongly interacting medium in the form of 
massive vector mesons).

\subsection{Pion form factor}

The electromagnetic form factors are important characteristics of stable had\-rons
and have been extensively studied experimentally.
These quantities encode information on the distribution of quarks and gluons within hadrons,
the knowledge of these distributions is crucial for understanding the transition of QCD from long to short distance scales.

Within the bottom-up holographic approach, a form factor of a state $\Phi$ is given by
a certain overlap integral in the fifth coordinate $z$ with external probe described by a bulk-to-boundary propagator~\cite{GR2007HW,Radyushkin2007,Brodsky2007}
(a short review and many references can be found in a recent Ref.~\cite{contreras2021}).
For instance, if $\Phi_0(z)$ is the holographic wave function of ground scalar state in the standard SW model with background $e^{cz^2}$,
the space-like electromagnetic form factors of this state (i.e., when the corresponding particle is probed by a virtual photon) is given by~\cite{br3}
\begin{equation}
\label{ff}
F(Q^2) = e \int\limits_0^\infty\frac{dz}{z^3}\,e^{cz^2}V(Q^2,z)\Phi_0^2(z),
\end{equation}
where $Q$ is the Euclidean momentum, the constant $e$ must be chosen such that $F(0)=1$ to satisfy
the charge conservation in Minkowski space at $Q^2=0$
(or, alternatively, this can be achieved by appropriate normalization of $\Phi_0(z)$),
and $V(Q^2,z)$ is the vector bulk-to-boundary propagator~\eqref{btb}.
In other words, the form factor in AdS is represented as the overlap in the holographic coordinate of the normalizable
modes dual to the incoming and outgoing hadrons with the non-normalizable mode dual to the external source.
The weight factor $1/z^3$ originates from the contraction $\sqrt{g}\,\varepsilon^M\partial_M\sim1/z^3$, where $\varepsilon^M$
is the polarization vector of external photon (with the condition~\eqref{18}, $\varepsilon^z=0$) and $\partial_M$ stems
from the Noether scalar current.

The expression for $\Phi_0(z)$ follows from the wave function~\eqref{wf2},
\begin{equation}
\label{phi}
\Phi_0(z)\sim e^{-(c+|c|)z^2/2} z^{2+|s|-J},
\end{equation}
with $|s|$ given by~\eqref{ss}. This function depends on the mass of corresponding mode only via $|s|$. If the constant part
in the effective mass~\eqref{meff2} is fixed, $\Phi_0(z)$ is not changed if we perform a general shift in the spectrum
with the help of $\mathcal{O}(z^2)$ contribution in~\eqref{meff2}. In principle, we can equate the mass of the ground state to zero
and try to interpret $\Phi_0(z)$ for $J=0$ as the holographic wave function of pion. The expression~\eqref{ff} has then the interpretation of
electromagnetic pion form factor. Substituting~\eqref{btb} and~\eqref{phi} into~\eqref{ff} we get
\begin{equation}
\label{ff2}
F(Q^2) = e \Gamma\lb 1+ \tilde{Q}^2\rb \int\limits_0^\infty dz\,z^{1+2(|s|-J)}e^{-(3|c|+c)z^2/2} U\lb  \tilde{Q}^2,0,|c|z^2\rb.
\end{equation}
Since $|s|\geq J$ this expression is convergent for any sign of $c$.
If the pion corresponded to twist-two QCD operator, we would have $|s|=J$ (see a comment after Eq.~\eqref{ss}).
The expression~\eqref{ff2} can be then integrated completely for $c<0$, i.e. in the case of standard SW$^-$ holographic model,
as shown analytically in Ref.~\cite{Radyushkin2007}.
Taking $e=2|c|$ to provide $F(0)=1$ and introducing the variable
$$y=|c|z^2,$$
the electromagnetic pion form factor takes the form
\begin{equation}
\label{ff3}
F_\pi(Q^2) = \Gamma\lb 1+ \tilde{Q}^2\rb \int\limits_0^\infty dy\, e^{-y} U\lb  \tilde{Q}^2,0,y\rb = \frac{1}{1+\tilde{Q}^2},
\end{equation}
remarkably reproducing the result of exact Vector Meson Dominance (VMD): $4|c|$ is the mass of the lightest vector meson in the standard SW model,
(see the relation~\eqref{polM} and we recall the notation~\eqref{tildeQ}, $\tilde{Q}^2\equiv\frac{Q^2}{4|c|}$).

In the LF holographic approach, one obtains the result~\eqref{ff3} due to the imposed condition~\eqref{LFm} which means $|s|=L=0$ for pions.
But as we commented in the previous section, there is no solution with discrete spectrum for $L=0$.

Within the standard AdS/QCD approach, we should take $\Delta=3$ for the pion states because it is the canonical dimension of the scalar quark bilinear
operator in QCD. This leads to $|s|=1$ for scalars and the result~\eqref{ff3} does not take place
(scales as $1/\tilde{Q}^4$~\cite{br3,Brodsky2007}) if we act in a straightforward manner.
This should mean that we do something wrong when associating $\Phi_0(z)$ with physical pion. Our error seems to lie in neglecting the
fact that the straightforward non-normalizable solution $\phi(z)$ of scalar equation of motion for $\Delta=3$ has the behavior $\phi(z)\sim z$ near the
AdS boundary, this also follows from the general UV asymptotics~\eqref{25a}.
For this reason, the physical source for scalar particles, which is finite at the UV boundary, must be associated
with $\phi(z)/z$ as is always accepted in holographic calculations of relevant scalar correlation functions. But this also imply that the
physical pion state should be associated\footnote{Remotely similar arguments (but leading to a more complex identification of the pion)
were used in Ref.~\cite{Radyushkin2007} for a holographic calculation of the pion form factor.} with $\Phi_\pi(z)=\Phi_0(z)/z$, hence, it is $\Phi_\pi(z)$ that
should be substituted into~\eqref{ff} in place of $\Phi_0(z)$. Technically this is tantamount to taking $|s|=0$ in~\eqref{phi} and the result~\eqref{ff3}
is reproduced.

Perhaps a more physical argument consists in interpreting the pion as the longitudinal component of the axial-vector field $A_{||} (z)$,
as is usual in the chiral effective field theories and in many AdS/QCD models starting from Refs.~\cite{son1,pom}.
Then indeed $\Phi_0(z)\sim z^2$ for the pion as it arises from a component of vector field corresponding to a
twist-two QCD operator, $A_{||} (z)\sim z^2$, but the pion mass does not follow from an equation of motion for scalar particles.
In the usual descriptions of spontaneous CSB, the fields $\Phi_\pi(z)$ and $A_{||} (z)$ are mixed, the details of mixing are not important for us
since the main property is the resulting scaling $\mathcal{O}(z^2)$, all arising factors will be absorbed into the normalization $F_\pi(0)=1$.

Another phenomenological difficulty encountered in the LF holographic approach is that the poles of vector bulk-to-boundary propagator~\eqref{repr}
(interpreted as ``dressed vector current'') do not match the LF vector spectrum~\eqref{LF} for the $S$-wave mesons, $L=0$. The experimental data on $F_\pi(Q^2)$
are well described for the vector spectrum with the behavior of radially excited states~\cite{Brodsky2007},
\begin{equation}
\label{rho}
m_\rho^2(n)\sim n+1/2,
\end{equation}
as in~\eqref{LF} or~\eqref{12lin}. As a result, the vector meson masses must be shifted manually to
their physical location to obtain a good agreement with data~\cite{br3}.

Let us demonstrate how one can proceed within the present approach. Introduction of non-zero intercept parameter $b$ in the spectra
with the help of $\mathcal{O}(z^2)$ contribution in~\eqref{meff2}, as was mentioned above, does not change the holographic pion wave function.
In the bulk-to-boundary propagator~\eqref{btb}, this leads to the shift $\tilde{Q}^2\rightarrow \tilde{Q}^2+b$ and one gets~\eqref{vc4}.
Consequently, the expression for the form factor can be readily obtained from~\eqref{ff3} by doing the same shift and
appropriately choosing the constant $e$ (to provide $F(0)=1$),
\begin{equation}
\label{ff4}
F_\pi(Q^2)  = \frac{1+b}{1+b+\tilde{Q}^2}=\frac{1}{1+\frac{Q^2}{4|c|(1+b)}}.
\end{equation}
The form factor~\eqref{ff4} can be compared with existing experimental data
on the space-like pion form factor at different choices of parameter $b$ and an optimal fit can be found.
The best fit is close to $b=-1/2$, see Fig.~\ref{ff_exp_plot} below.
Since the radially excited vector spectrum of SW model behaves as
\begin{equation}
\label{rho2}
m_\rho^2(n)\sim n+1+b,
\end{equation}
the value of
\begin{equation}
\label{b12}
b=-\frac12,
\end{equation}
for the intercept parameter results in the behavior~\eqref{rho}.

The cases $c=0$ but formally $|c|\neq0$ and $c>0$ in~\eqref{ff2} correspond to the SW$^0$ (``No-wall'') and SW$^+$ holographic models.
The form factor~\eqref{ff3} for these models can be compactly written as
\begin{equation}
\label{ff3c}
F_\pi(Q^2) = e(k)\Gamma\lb 1+ \tilde{Q}^2\rb \int\limits_0^\infty dy\, e^{-(1+k)y} U\lb  \tilde{Q}^2,0,y\rb,
\end{equation}
\begin{equation}
\text{SW}^+\text{:}\quad k=1;\qquad\qquad \text{SW}^0\text{:}\quad k=\frac12.
\end{equation}

The integral in~\eqref{ff3c} at $k\neq0$ can be solved with the help of the same trick as was
used in Ref.~\cite{Radyushkin2007} to derive~\eqref{ff3} and~\eqref{repr}.
The trick consists in exploiting a known integral representation of the Tricomi function~\cite{res}
to rewrite the bulk-to-boundary propagator~\eqref{btb} of SW$^-$ model (i.e., $c<0$) as
\begin{equation}
\label{btb2}
  V^-(\tilde{Q}^2,y)= y\int\limits_0^1 \frac{x^{\tilde{Q}^2}dx}{(1-x)^2}e^{-\frac{yx}{1-x}}.
\end{equation}
Substituting this representation into~\eqref{ff} for any sign of $c$ and choosing the pion wave function as before, we get
\begin{multline}
F_\pi(Q^2) = e(k)\int\limits_0^1\frac{x^{\tilde{Q}^2}dx}{(1-x)^2} \int\limits_0^\infty dy\,y\, e^{-\left(\frac{1}{1-x}+k\right)y}=\\
e(k)\int\limits_0^1\frac{x^{\tilde{Q}^2}dx}{(1-x)^2}\left(\frac{1}{1-x}+k\right)^{-2}=
e(k)\int\limits_0^1 \frac{x^{\tilde{Q}^2}dx}{\left(1+k-kx\right)^2}.
\end{multline}
For $k=0$ the integral is elementary and one arrives at the known relation~\eqref{ff3}.
At $k\neq0$ the emerged integral cannot be solved in terms of elementary functions
but can be expressed through the  Lerch transcendent function $\Phi(t,s,\xi)$~\cite{res}.
Setting $e(k)=k+1$ to satisfy $F_\pi(0)=1$, we obtain the final answer for the normalized pion form factor,
\begin{equation}
\label{ff6}
F_\pi(Q^2) = 1+\frac{1}{k}-\frac{\tilde{Q}^2}{k}\,\Phi\left(\frac{k}{1+k},1,\tilde{Q}^2\right).
\end{equation}
An important caveat: The normalization $F_\pi(0)=1$ is understood as the limit,
\begin{equation}
\lim\limits_{Q^2\rightarrow +0}F_\pi(Q^2)=1,
\end{equation}
because if we set $Q^2=0$ in~\eqref{ff6} we would obtain $F_\pi(0)=1+1/k$. This mathematical subtlety of Lerch function $\Phi$
must be taken into account --- we make use of the physical limit at zero Euclidean momentum.

It is interesting to notice in passing that the appearance of Lerch transcendent function looks intriguing since this function generalizes
the polylogarithm, $\text{Li}_s(t)=t\Phi(t,s,1)$, that at positive integer $s$ arises in the calculations of higher-order Feynman diagrams
in quantum electrodynamics.
Within the framework of SW holographic models, this function emerges in gravitational form factors of nucleons~\cite{zahed}.

One can write an alternative to~\eqref{ff6} expression in a form of convergent for $k>0$ series in terms of Euler digamma and Beta functions,
\begin{equation}
\label{ff6c}
F_\pi(Q^2) = 1+\sum_{n=0}^{\infty}\frac{\psi\left(\tilde{Q}^2+n+1\right)-\psi\left(n+1\right)-\ln (1+k)}{(1+k)^{n+1}B\left(\tilde{Q}^2,n+1\right)}.
\end{equation}

\begin{figure}[!ht]
 \center{\includegraphics[width=0.7\linewidth]{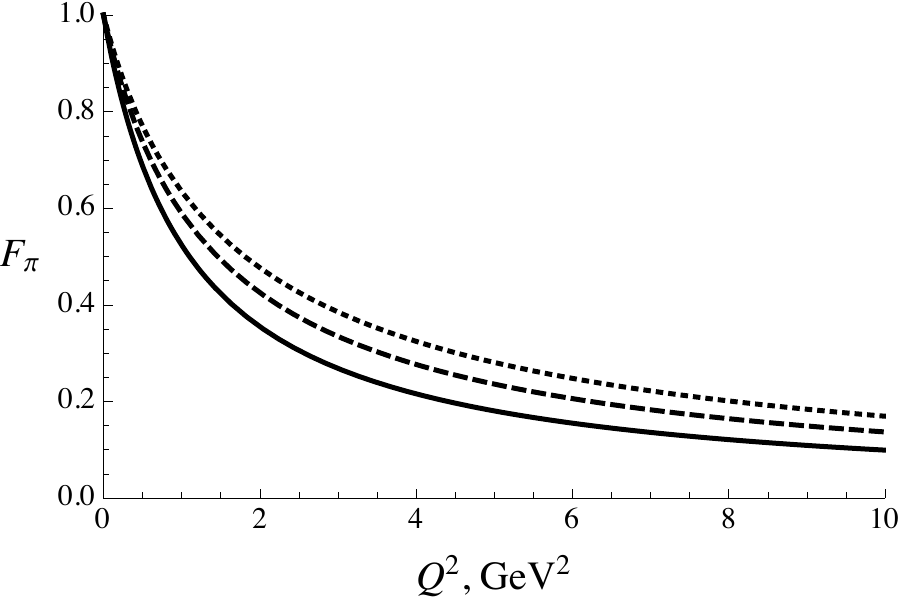}}
 \caption{\small The pion form factors~\eqref{ff3} (solid) and~\eqref{ff6} for $k=1/2$ (dashed) and $k=1$ (dotted).
 The scale parameter $4|c|=1.1\,\text{GeV}^2$, as in Fig.~\ref{ff_exp_plot}.}
 \label{ff_plot}
\end{figure}
The plots for~\eqref{ff3} and~\eqref{ff6} are displayed in Fig.~\ref{ff_plot}. It is qualitatively seen that enlarging $k$
we get progressively slower fall-off with Euclidean momentum.
This happens due to a contribution of radially excited vector mesons.
The given effect can be analytically visualized using the definition of function $\Phi(t,s,\xi)$~\cite{res},
\begin{equation}
\label{phidef}
\Phi(t,s,\xi)=\sum_{n=0}^\infty\frac{t^n}{(n+\xi)^s},
\end{equation}
in the expression~\eqref{ff6},
\begin{equation}
\label{ff7}
F_\pi(Q^2) = 1-\frac{\tilde{Q}^2}{k}\sum_{n=1}^\infty\left(\frac{k}{1+k}\right)^n\frac{1}{\tilde{Q}^2+n},
\end{equation}
where the $n=0$ contribution canceled the term $1/k$ in~\eqref{ff6}. The obtained representation can be
further simplified writing $\tilde{Q}^2=\tilde{Q}^2+n-n$ in the numerator and making use of the summation
\begin{equation}
\sum_{n=1}^\infty\left(\frac{k}{1+k}\right)^n=k,
\end{equation}
that results in the following pole expansion,
\begin{equation}
\label{ff8}
F_\pi(Q^2) = \frac{1}{k}\sum_{n=1}^\infty\left(\frac{k}{1+k}\right)^n\frac{n}{\tilde{Q}^2+n}.
\end{equation}
The obtained relation shows explicitly how the contributions of highly excited states are damped depending on a model.
In the SW$^+$ model, $k=1$, the damping factor is $\frac{1}{2^n}$. In the No-Wall model, $k=\frac12$, the damping factor
becomes stronger, $\frac{1}{3^n}$. The SW$^-$ model, $k=0$, corresponds to the extreme case of infinite damping
when only the first term survives leading to the well known result~\eqref{ff3}.

An alternative way to get the results above is to substitute the representation for the bulk-to-boundary propagator~\eqref{repr}
to the holographic definition of pion form factor~\eqref{ff} and use the following property of the Laguerre polynomials~\cite{res},
\begin{equation}
\int\limits_0^\infty dy\,y\, e^{-y}L_n^1(y)=\delta_{n0}.
\end{equation}
This relation is responsible for the absence of contribution of higher modes, $n>0$, in the SW$^-$ model, thus realizing the exact VMD.
In the case of SW$^+$ and SW$^0$ models, the exponent is replaced by $e^{-(1+k)y}$ that invalidates
the given relation for $k\neq0$. The use of Lerch transcendent function, however, looks nicer since it directly yields
the rate of damping of contribution from the excited states.

The nonzero intercept parameter $b$ can be simply introduced in the form factor~\eqref{ff6} via the shift $\tilde{Q}^2\rightarrow \tilde{Q}^2+b$,
as before. This changes the normalization factor, $e = e(k,b)$ (that was previously normalized as $e=e(k,0)=1$), the final result is
\begin{equation}
\label{ff6b}
F_\pi(Q^2) = e(k,b)\left[1+\frac{1}{k}-\frac{\tilde{Q}^2+b}{k}\,\Phi\left(\frac{k}{1+k},1,\tilde{Q}^2+b\right)\right],
\end{equation}
\begin{equation}
\label{ff6b2}
e(k,b)=\left[1+\frac{1}{k}-\frac{b}{k}\,\Phi\left(\frac{k}{1+k},1,b\right)\right]^{-1}.
\end{equation}

Now one can try to get an optimal fit for $b$ by comparing the relation~\eqref{ff6b} with the experimental data for $F_\pi(Q^2)$ in the space like region.
A couple of examples is displayed in Fig.~\ref{ff_exp_plot}. The mean slope $4|c|$ of Regge and radial Regge trajectories
in the spectra of light non-strange mesons was fitted in reviews~\cite{bugg,phen4}, the result was $4|c|\approx1.1$~GeV$^2$
(see a brief review in the Appendix~D). We use everywhere this input value for the slope of meson trajectories.
\begin{figure}[!ht]
 \center{\includegraphics[width=0.9\linewidth]{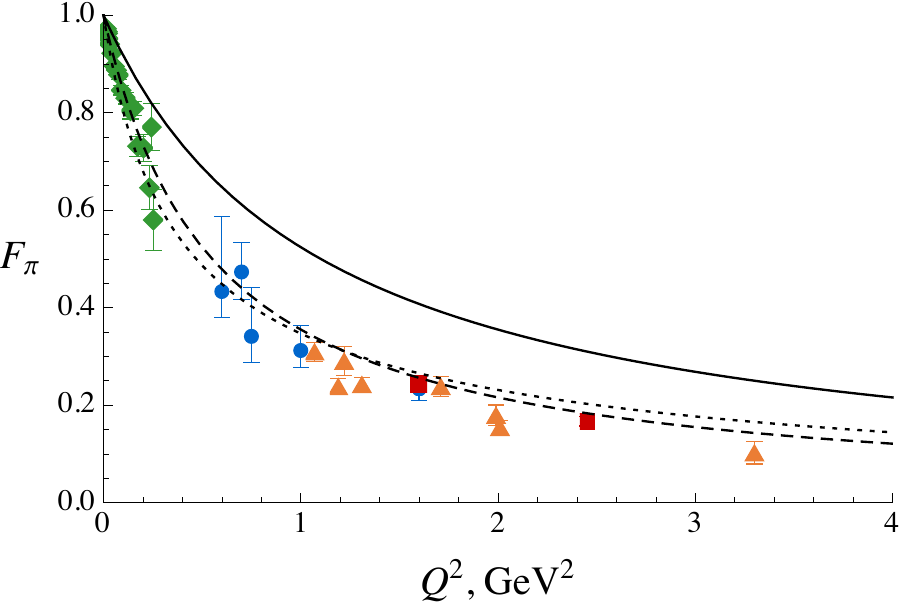}}
 \caption{\small The pion form factor~\eqref{ff4}: $b=0$ (solid) and $b=-1/2$ (dashed).
 The dotted line depicts the form factor~\eqref{ff6b} for
 $k=1$ and $b\approx-0.7324$. Everywhere the scale parameter $|c|$ is fixed from the
 phenomenological slope $4|c|=1.1$~GeV$^2$ (see the text). The experimental data is taken
 from: diamonds --- Ref.~\cite{ff0}, circles --- Ref.~\cite{ff060},
 squares --- Ref.~\cite{ff160}, triangles --- data compiled in Ref.~\cite{ffcomp}.}
 \label{ff_exp_plot}
\end{figure}

In the case of SW$^-$ model, the best fit is achieved at
\begin{equation}
\label{b}
b^-\approx-0.51.
\end{equation}
This value perfectly matches the expected pattern of CSB in the light meson spectrum mentioned after Eq.~\eqref{J0}
and agrees with the previous fit~\eqref{b12} for the SW$^-$ model.
The global fit~\eqref{hydr} of Regge spectrum in the light non-strange mesons also resulted in a mean
intercept value close to~\eqref{b}.
It is also interesting to note that the numerical solution $b_1^+$ in~\eqref{eqb3} providing a self-consistency of SW$^+$ model
practically coincides with~\eqref{b}.

In the case of SW$^+$ models, the best fit is achieved at $b^+\approx-0.73$.
For the best fits, the curves in Fig.~\ref{ff_exp_plot} lie very close. In order to estimate a variation of $b$ we show
in Fig.~\ref{ff_exp_plot} a curve for the SW$^+$ model that provides the best fit of data in the interval
$0.5\lesssim Q^2 \lesssim 2.5$ GeV$^2$. This curve corresponds to $b^+\approx-0.77$ (for the SW$^-$ model, such a
fitting would give $b^-\approx-0.58$). Unfortunately, with such small intercept the SW$^+$ model predicts the mass of the
first $\rho$-meson in~\eqref{rho2} near $0.5$~GeV, somewhat away from the experimental value of $0.77$~GeV~\cite{pdg}.

We observe thus that the SW$^-$ describes both the spectroscopy and pion form factor with a universal intercept $b^-\approx-1/2$,
while the SW$^+$ is not so successful. The situation with fits in the SW$^0$ model lies in between. Recalling that the SW$^-$ model
reproduces the exact VMD, we just rediscover once more an old phenomenological wisdom about the VMD concept.

The physical value of intercept $b$ can be independently estimated
from the mean pion charge radius squared. This quantity appears in the small $Q^2$ expansion of the electromagnetic
pion form factor~\cite{pdg},
\begin{equation}
\label{ff5}
F_\pi(Q^2)  = 1-\frac16\langle r^2_\pi\rangle Q^2+\mathcal{O}(Q^4).
\end{equation}
This relation should be matched to the small $Q^2$ expansion of our theoretical prediction~\eqref{ff6b}.

As a warm-up exercise, let us first match~\eqref{ff5} to the small $Q^2$ expansion of~\eqref{ff4}. The result
can be written immediately,
\begin{equation}
\label{radius}
\langle r^2_\pi\rangle=\frac{6}{4|c|(1+b)}.
\end{equation}
Experimentally $\langle r^2_\pi\rangle=0.659^2$ fm$^2$ with a high precision~\cite{pdg}.
Using our value $4|c|=1.1$~GeV$^2$ for the slope and the conversion factor
$1\,\text{fm}=(0.197\,\text{GeV})^{-1}$ we obtain exactly the estimate~\eqref{b}.
The correct prediction of $\langle r^2_\pi\rangle$ means the correct prediction of the experimental slope of
pion form factor at small $Q^2$.
So the overall phenomenological consistency of the SW$^-$ model looks almost perfect.

For the small $Q^2$ expansion of the full answer~\eqref{ff6b} one can use an identity~\cite{res}
$$\partial_\xi\Phi(t,s,\xi)=-s\Phi(t,s+1,\xi)$$
that leads to the following expression for $\langle r^2_\pi\rangle$,
\begin{equation}
\label{radius2}
\langle r^2_\pi\rangle=\frac{6}{4|c| e(k,b) k}\left[\Phi\left(\frac{k}{1+k},1,b\right)-b\Phi\left(\frac{k}{1+k},2,b\right)\right].
\end{equation}
It should be remarked that the limits $k\rightarrow0$ and $b\rightarrow0$ are not completely interchangeable in~\eqref{radius2} .
The first limit can be taken using a series representation for the Lerch transcendent function~\cite{res},
\begin{equation}
\Phi(t,s,\xi)=\frac{1}{1-t}\sum_{n=0}^\infty\left(\frac{-t}{1-t}\right)^n\sum_{i=0}^n(-1)^i\binom{n}i(\xi+i)^{-s},
\end{equation}
from which we get
\begin{equation}
\Phi\left(\frac{k}{1+k},s,b\right)_{k\rightarrow0}=\frac{1}{b^s}+\frac{k}{(1+b)^s}+\mathcal{O}(k^2),
\end{equation}
for $s=1$ and $s=2$. Making use of this expansion we reproduce both $e(0,b)=(1+b)^{-1}$ in~\eqref{ff4} from~\eqref{ff6b2}
and~\eqref{radius} from~\eqref{radius2}, i.e. the transition to the SW$^-$ model is analytically smooth
(but this is not the case for all software systems).

Substituting to the Eq.~\eqref{radius2} the experimental value of $\langle r^2_\pi\rangle$ and our phenomenological value
for the slope $4|c|$, we obtain an equation for the intercept $b$ which can be solved numerically. For the SW$^+$ model,
$k=1$, the result is $b^+\approx-0.64$; for the No-Wall model, $k=1/2$, we get $b^0\approx-0.59$, lying as expected between $b^+$
and $b^-$ in~\eqref{b}. The obtained estimates are noticeably closer to $b=-1/2$ than the estimates above based on a global
fit to the experimental data on $F_\pi(Q^2)$.

\section{Conclusions}

We have substantially advanced in construction of a general bottom-up holographic theory of linear meson Regge trajectories.
Many Soft Wall (SW) like holographic models and their relatives proposed in the literature can be
obtained as particular cases within our framework.
The proposed theory is based on the most general quadratic in fields action in the AdS$_5$ background
in which the Poincar\'{e} invariance along the fifth coordinate is violated in such a way that
the mass spectrum of normalizable 4D modes has the Regge form. The form of this action is not
unique and we scrutinized various interrelations between different forms. Perhaps the simplest form
arises when the action is written in the pure AdS$_5$ space without any $z$-dependent background
(here $z$ is the holographic coordinate). In this case, the 5D mass term contains $\mathcal{O}(z^2)$
and $\mathcal{O}(z^4)$ contributions which are responsible for the intercept and slope of
linear Regge trajectories, correspondingly. We argued that these contributions describe the effects
caused by the most important non-local phenomena of strongly coupled 4D gauge theory ---
the confinement and spontaneous chiral symmetry breaking --- in terms of a local 5D dual field theory
formulated in the AdS$_5$ space. In addition, the most general quadratic in fields holographic action
can contain a $z$-dependent contribution of another sort, namely which is linear in derivative $\partial_z$ of a field.
This contribution regulates the ratio of slopes of radial and angular trajectories and,
to the best of our knowledge, was never explored before.

We performed a detailed comparison of the proposed holographic theory with some other approaches
describing the linear Regge trajectories --- the Light Front holographic QCD, Veneziano like dual amplitudes, and
string like phenomenological spectra.

We considered several application of the constructed holographic theory.
The first application was a study of confining behavior using a certain formulation of
holographic Wilson criterion. It was found that this criterion can be satisfied
for the spin-one and spin-two particles. These two cases have interesting massless limits.

As the second application, some effects of chiral symmetry breaking in the meson spectra were discussed.
We drew attention to the fact that this phenomenon seems to affect the whole leading trajectories,
not just the low-spin states as in the standard descriptions.
A possible direction for a holographic modeling of this phenomenon was outlined.

The third considered application concerned the vector two-point correlator. This correlator has a notorious
problem of unphysical massless pole which is absent only in the special case of simple SW holographic model
with negative exponential background that was introduced
in the original paper~\cite{son2}. We constructed a general recipe how to avoid this problem
if a different form or an extended version of SW model is used. The requirement of absence of
massless pole is converted into a prediction of intercept of the radial Regge trajectory. The prediction
turns out to be different for different forms of SW model. 

The fourth application was devoted to the electromagnetic form factor of charged pion.
We highlighted differences between the results produced by various forms of holographic
SW models. Most notably, the exact vector meson dominance takes place only in the case of the SW model
with negative exponential background, within the framework of other formulations,
all radially excited states contribute. We derived a model-dependent damping rate of the
contribution of radial excitations.

Let us outline other possible applications and developments.
Following the calculation of Ref.~\cite{AZ}, the results of Section 5.1 (where consistent corrections to the AdS metrics
within the constructed general theory were discussed) can be used to extract the precise forms of heavy-quark potentials.
Also the models in this form can be explored for analysis of holographic thermodynamics.
Our general discussion of holographic modeling of chiral symmetry breaking should
be further elaborated and concrete models should be constructed. Concerning the hadron structure, many problems can be addressed ---
the anomalous form factor of neutral pion, the electromagnetic and gravitational nucleon form factors together with related problems
of various generalized quark and gluon distributions and with electroproduction of heavy vector mesons.

Finally we should mention that the linearity of radial Regge trajectories guarantees the expected analytical form of
Operator Product Expansion (the parton logarithm plus polynomial in inverse Euclidean momentum squared corrections) of two-point
correlation functions. The required analytical properties of correlators are spoiled immediately if nonlinear corrections
are introduced, both manually and via a back-reacted geometry. Whether the proposed theory can be extended to non-linear case
in a self-consistent way is an open problem.

\section*{Acknowledgments}

This research was funded by the Russian Science Foundation grant number 21-12-00020.


\newpage

\appendix

\section*{Appendix A}
\addcontentsline{toc}{abcd}{\bf Appendix A}

\renewcommand\theequation{A.\arabic{equation}}
\setcounter{equation}{0}

Consider the following action for tensor fields,
\begin{equation}
\label{ap1}
  S=\frac{1}{2}\int d^5x\sqrt{g}e^{cz^2}
  g^{MN}g^{M_1N_1}\dots g^{M_JN_J}\partial_M \Phi_{M_1\dots M_J}\partial_N\Phi_{N_1\dots N_J}.
\end{equation}
The equation of motion with the condition~\eqref{18} is
\begin{equation}
  \lsb \partial_\mu\partial^\mu-z^{3-2J}e^{-cz^2}\pz\lb z^{2J-3}e^{cz^2}\pz\rb\rsb\Phi^{\mu_1\dots \mu_J}=0,
\end{equation}
We set $R=1$ in this Appendix.
Within the plane wave ansatz~\eqref{39} for the discrete spectrum of 4D particles,
the equation takes the form
\begin{equation}
  z^{2J-3}e^{cz^2}m_n^2v^{(J)}_n+\pz\lb z^{2J-3}e^{cz^2}\pz v^{(J)}_n\rb=0.
\end{equation}
Making the substitution
\begin{equation}
\label{sub2}
v^{(J)}_n=z^{(3-2J)/2}e^{-cz^2/2}\psi_n,
\end{equation}
we get after a straightforward algebra
\begin{equation}
\label{sr1}
  -\psi''_n+\lsb c^2z^2+2c(J-1)+\frac{(J-2)^2-1/4}{z^2}\rsb\psi_n=m_n^2\psi_n.
\end{equation}
After adding the mass term $m_5^2\Phi^{\dots}\Phi_{\dots}$ to~\eqref{ap1}, this equation
is extended to
\begin{equation}
\label{sr2}
  -\psi''_n+\lsb c^2z^2+2c(J-1)+\frac{(J-2)^2+m_5^2-1/4}{z^2}\rsb\psi_n=m_n^2\psi_n.
\end{equation}
This equation has a form of one-dimensional Schr\"{o}dinger equation
\begin{equation}
\label{sr3}
-\psi_n''+\left[x^2+\frac{s^2-1/4}{x^2}\right]\psi_n=E_n\psi_n,
\end{equation}
which arises in the quantum-mechanical problem of two-dimensional harmonic oscillator
with orbital momentum $s$. The corresponding spectrum of discrete modes is well known,
\begin{equation}
E_n=4n+2|s|+2, \qquad n=0,1,2,\dots, \quad |s|>1/2,
\end{equation}
with the normalized eigenfunctions
\begin{equation}
\label{wf}
\psi_n=\sqrt{\frac{2n!}{(|s|+n)!}}\,e^{-x^2/2}x^{|s|+1/2}L_n^{|s|}(x^2),
\end{equation}
where $L_n^{|s|}(x^2)$ are associated Laguerre polynomials. The first two polynomials are
\begin{equation}
L_0^{|s|}(x^2)=1,\qquad L_1^{|s|}(x^2)=1+|s|-x^2.
\end{equation}
This allows to write immediately the spectrum of equation~\eqref{sr2},
\begin{equation}
\label{primary}
m_{n,J}^2=2|c|\left(2n+1+\frac{c}{|c|}(J-1)+\sqrt{(J-2)^2+m_5^2}\right).
\end{equation}
The relations~\eqref{sub2} and~\eqref{wf} dictate the corresponding eigenfunctions,
\begin{equation}
\label{wf2}
v^{(J)}_n=
e^{-(c+|c|)z^2/2} z^{2+|s|-J} L_n^{|s|}(|c|z^2),
\end{equation}
where
\begin{equation}
\label{ss}
|s|=\sqrt{(J-2)^2+m_5^2},
\end{equation}
and $m_5^2$ is given by the relation~\eqref{3c}, $m_5^2=(\Delta-J)(\Delta+J-4)$.
In the case of dimensions of twist-two operators, $\Delta=J+2$, we obtain $|s|=J$.

The incorporation of non-zero intercept normalized as in the text (e.g., as in Eq.~\eqref{vect3})
is equivalent to adding the constant $4b|c|$ to the potential of Schr\"{o}dinger equation~\eqref{sr2}.
This does not change the eigenfunction~\eqref{wf2}.
For $|s|=J$ the spectrum~\eqref{primary} becomes
\begin{equation}
\label{primary2}
m_{n,J}^2=4|c|\left[n+\frac12\left(J+1+\frac{c}{|c|}(J-1)\right)+b\right].
\end{equation}

\section*{Appendix B}
\addcontentsline{toc}{abcd}{\bf Appendix B}

\renewcommand\theequation{B.\arabic{equation}}
\setcounter{equation}{0}

In this Appendix, we provide a more rigorous version of the analysis of Section~4.1 and extend this analysis
to the tensor case using the generalized background derived in Section~4.2.

Consider the following 5D action,
\begin{equation}
\label{45}
S=\frac{1}{2}\int d^5 x \sqrt{g}e^{-cz^2}U^2(b,0,cz^2)g^{MR}g^{NS}\partial_M V_N \partial_R V_S,
\end{equation}
where $U$ is the Tricomi confluent hypergeometric function and a general normalization factor is omitted.
Actually the antisymmetric tensor field
was discussed in~\cite{genSW} but the SW model in the form written in~\eqref{45} is more convenient for
further generalization to symmetric tensor fields and to scalar fields. The result does not depend on the chosen form,
but for completeness below we will repeat our analysis for the antisymmetric case.

Let us eliminate the ``generalized'' background with the help of appropriate field redefinition,
\begin{equation}
  V_M\equiv C v_M, \quad C\equiv e^{cz^2/2}U^{-1}(b,0,cz^2).
\end{equation}
In what follows, we omit the arguments of the Tricomi function, since they are always the
same in our consideration. This transformation changes the derivative with respect to the holographic coordinate \(z\),
\begin{equation}
  \pz V_M= cz e^{cz^2/2}U^{-1}v_M-e^{cz^2/2}U^{-2}U'2czv_M+e^{cz^2/2}U^{-1}\pz v_M,
\end{equation}
where \(U'\) denotes the derivative with respect to the third argument.
Let us introduce another notation,
\begin{equation}
  G\equiv \lb 1-2\frac{U'}{U}\rb cz,
\end{equation}
which we use to rewrite the field derivative as
\begin{equation}
  \pz V_M\equiv C\pz v_M+CGv_M.
\end{equation}
Consider in detail the kinetic term,
\begin{multline}
g^{MR}g^{NS}\partial_M V_N \partial_R V_S=
    g^{\mu\rho}g^{NS}\partial_\mu V_N \partial_\rho V_S+g^{zz}g^{NS}\pz V_N \pz V_S=\\
    C^2g^{\mu\rho}g^{NS}\partial_\mu v_N \partial_\rho v_S+
    g^{zz}g^{NS}\lb C\pz v_N+CGv_N\rb\lb C\pz v_S+CGv_S\rb=\\
    C^2g^{MR}g^{NS}\partial_M v_N \partial_R v_S+
    C^2g^{zz}g^{NS}\lb 2Gv_N\pz v_S+G^2v_Nv_S\rb.
\end{multline}
Substituting this expression back into the action, the factor \(C^2\) is canceled out,
\begin{equation}
  S=\frac{1}{2}\int d^5 x\sqrt{g}\lsb g^{MR}g^{NS}\partial_M v_N \partial_R v_S+
  g^{zz}g^{NS}\lb 2Gv_N\pz v_S+G^2v_Nv_S\rb\rsb.
\end{equation}
The variation of the action reads
\begin{equation}
  \begin{aligned}
    \delta_v S&=\int d^5 x\sqrt{g}\lsb
    g^{MR}g^{NS}\partial_M v_N \partial_R \delta v_S+g^{zz}g^{NS}G\delta v_N\pz v_S+\right.\\&+\left.
    g^{zz}g^{NS}Gv_N\pz \delta v_S+g^{zz}g^{NS}G^2v_N\delta v_S\rsb.
  \end{aligned}
\end{equation}
After integrating by parts,
\begin{equation}
  \begin{aligned}
    \delta_v S&=\int d^5 x\lsb
    -\partial_R\lb\sqrt{g}g^{MR}g^{NS}\partial_M v_N\rb+
    \sqrt{g}g^{zz}g^{NS}G\pz v_N-\right.\\&-\left.
    \pz\lb\sqrt{g}g^{zz}g^{NS}Gv_N\rb+\sqrt{g}g^{zz}g^{NS}G^2v_N\rsb\delta v_S,
  \end{aligned}
\end{equation}
a part of the third term (with derivative acting on the field) cancels the second term, the resulting variation is
\begin{multline}
  \delta_v S=\frac{1}{2}\int d^5 x\lsb
  -\partial_R\lb\sqrt{g}g^{MR}g^{NS}\partial_M v_N\rb-\right.\\\left.-
  \lb \pz\lb\sqrt{g}g^{zz}g^{NS}G\rb -
  \sqrt{g}g^{zz}g^{NS}G^2\rb v_N\rsb\delta v_S,
\end{multline}
from which the equation of motion follows
\begin{equation}
  \lsb-\partial_R\lb\sqrt{g}g^{MR}g^{NS}\partial_M \rb-
  \pz\lb\sqrt{g}g^{zz}g^{NS}G\rb +
  \sqrt{g}g^{zz}g^{NS}G^2\rsb v_N=0.
\end{equation}
After substituting various metric-related factors,
\begin{equation}
  \sqrt{g}g^{MR}g^{NS}=\frac{R}{z}\eta^{MR}\eta^{NS},\quad
  \sqrt{g}g^{zz}g^{NS}=-\frac{R}{z}\eta^{NS},
\end{equation}
and multiply by \(z/R\) we get
\begin{equation}
\label{v_simple_simplified_eom}
  \lsb-\partial_\mu\partial^\mu+
  z\pz\lb\frac{\pz}{z}\rb+
  z\pz\lb\frac{G}{z}\rb-
  G^2\rsb\eta^{NS}v_N=0.
\end{equation}
Let us expand the second $z$-derivative,
\begin{equation}
\label{gensw_v_eom}
  \lsb-\partial_\mu\partial^\mu+
  z\pz\lb\frac{\pz}{z}\rb+
  \pz G-\frac{G}{z}-G^2\rsb\eta^{NS}v_N=0.
\end{equation}
and examine the last three terms separately. The first term reads
\begin{equation}
  \pz G = c-2c\pz\lb\frac{zU'}{U}\rb=
  c-2c\lsb\frac{U'}{U}+2cz^2\frac{U''}{U}-2cz^2\frac{U'^2}{U^2}\rsb.
\end{equation}
The Tricomi function $U(b,k,x)$ is a solution of the differential equation~\eqref{kummer}
\begin{equation}
\label{KumEq}
  x\frac{d^2U}{dx^2}+(k-x)\frac{dU}{dx}=bU.
\end{equation}
In our case of $k=0$ and $x=cz^2$, this means
\begin{equation}
\label{3.18}
  cz^2U''-cz^2U'=bU,
\end{equation}
and we can write
\begin{equation}
\label{t1}
  \pz G =
    c-(2c+4c^2z^2)\frac{U'}{U}-4cb+4c^2z^2\frac{U'^2}{U^2}.
\end{equation}
The last two terms are
\begin{equation}
\label{t2}
  \frac{G}{z}=
  c-2c\frac{U'}{U},
\end{equation}
\begin{equation}
\label{t3}
  G^2=
  c^2z^2-4c^2z^2\frac{U'}{U}+4c^2z^2\frac{U'^2}{U^2}.
\end{equation}
Substituting the expressions \eqref{t1}, \eqref{t2}, and \eqref{t3} into the equation~\eqref{gensw_v_eom},
it is easy to arrive at the final version of the equation of motion,
\begin{equation}
\lsb\partial_\mu\partial^\mu-
  z\pz\lb\frac{\pz}{z}\rb+
  c^2z^2+4cb\rsb\eta^{NS}v_N=0.
\end{equation}
This equation represents a particular case of Eq.~\eqref{38} with $J=1$ and $m_\text{eff}^2R^2=c^2z^4+4cbz^2$,
the spectrum of discrete normalizable modes is given by~\eqref{41} which for $\tilde{b}=4|\tilde{c}|b$ under consideration
is the spectrum~\eqref{vect} that we wanted to obtain. This proves our statement.

The analysis above can be simply extended to the case of antisymmetric tensor field
which was used in the generalized SW vector model of Ref.~\cite{genSW}.
Consider the 5D action
\begin{equation}
\label{gen_sw_action}
  S=-\frac{1}{4}\int d^5 x \sqrt{g}e^{-az^2}U^2(b,0,az^2)g^{MR}g^{NS}F_{MN}F_{RS},
\end{equation}
where $F_{MN}=\partial_M V_N-\partial_N V_M$ and we preserve the notations of Ref.~\cite{genSW}.
A general normalization factor is omitted.
Our goal is to eliminate the generalized SW background by doing a field substitution
\begin{equation}
  V_M\equiv C v_M, \quad C\equiv e^{az^2/2}U^{-1}(b,0,az^2).
\end{equation}
We also define the equivalent of the antisymmetric tensor for the transformed field \(v_M\)
\begin{equation}
  f_{MN}\equiv\partial_M v_N-\partial_N v_M.
\end{equation}
The \(z\)-derivative acts as (the arguments of the Tricomi function is omitted in what follows)
\begin{equation}
  \pz V_M=
  az e^{az^2/2}U^{-1}v_M - e^{az^2/2}U^{-2}U'2azv_M+e^{az^2/2}U^{-1}\pz v_M,
\end{equation}
where \(U'\) denotes the derivative with respect to the third argument.
For the sake of convenience we repeat the notation,
\begin{equation}
  G\equiv \lb 1-2\frac{U'}{U}\rb az,
\end{equation}
in terms of which the field derivative is
\begin{equation}
  \pz V_M\equiv C\pz v_M+CGv_M.
\end{equation}
The 5D derivative can be written as:
\begin{equation}
  \partial_M V_N\equiv C\partial_M v_N+CGv_N\delta_M^z.
\end{equation}
The integrand in the action is
\begin{equation}
  \begin{aligned}
    g^{MR}&g^{NS}F_{MN}F_{RS}=\\&=
    g^{MR}g^{NS}C^2\lb f_{MN}+Gv_N\delta_M^z-Gv_M\delta_N^z\rb
    \lb f_{RS}+Gv_S\delta_R^z-Gv_R\delta_S^z\rb=\\&=
    g^{MR}g^{NS}C^2\lsb f_{MN}f_{RS}+Gf_{MN}\lb v_S\delta_R^z-v_R\delta_S^z\rb
    +Gf_{RS}\lb v_N\delta_M^z-v_M\delta_N^z\rb+\right.\\&+\left.
    G^2\lb v_S\delta_R^z-v_R\delta_S^z\rb\lb v_N\delta_M^z-v_M\delta_N^z\rb\rsb.
  \end{aligned}
\end{equation}
The second and the third term in the square brackets are equal since we can rename contracted
indices \(M\leftrightarrow R\) and \(N\leftrightarrow S\),
\begin{equation}
  \begin{aligned}
    g^{MR}g^{NS}F_{MN}F_{RS}&=
    g^{MR}g^{NS}C^2\lsb f_{MN}f_{RS}+2Gf_{MN}\lb v_S\delta_R^z-v_R\delta_S^z\rb+\right.\\&+\left.
    G^2\lb v_S\delta_R^z-v_R\delta_S^z\rb\lb v_N\delta_M^z-v_M\delta_N^z\rb\rsb.
  \end{aligned}
\end{equation}
We can further simplify the second term using the antisymmetry of \(f_{MN}\) and
renaming indices again (we temporarily omit the \(2GC^2\) factor),
\begin{equation}
  \begin{aligned}
    g^{MR}&g^{NS}f_{MN}\lb v_S\delta_R^z-v_R\delta_S^z\rb=
    \\&=
    g^{MR}g^{NS}f_{MN}v_S\delta_R^z-g^{NS}g^{MR}f_{NM}v_S\delta_R^z=
    2g^{MR}g^{NS}f_{MN}v_S\delta_R^z=\\&=
    2g^{Mz}g^{NS}f_{MN}v_S=
    2g^{zz}g^{NS}f_{zN}v_S.
  \end{aligned}
\end{equation}
The last term can be rewritten as (now we omit the \(G^2C^2\) factor)
\begin{equation}
  \begin{aligned}
    g^{MR}&g^{NS}\lb v_S\delta_R^z-v_R\delta_S^z\rb\lb v_N\delta_M^z-v_M\delta_N^z\rb=\\&=
    g^{zz}g^{NS}v_Sv_N-g^{zR}g^{Nz}v_Rv_N-g^{Mz}g^{zS}v_Sv_M+g^{MR}g^{zz}v_Rv_M=\\&=
    2g^{zz}g^{MN}v_Mv_N-2g^{Mz}g^{Nz}v_Mv_N.
  \end{aligned}
\end{equation}
Combining all this together, we get the following transformation for the integrand
\begin{equation}
  \begin{aligned}
    g^{MR}g^{NS}F_{MN}F_{RS}&=
    C^2\lsb g^{MR}g^{NS}f_{MN}f_{RS}+4Gg^{zz}g^{NS}f_{zN}v_S+\right.\\&+\left.
    2G^2g^{zz}g^{MN}v_Mv_N-2G^2g^{Mz}g^{Nz}v_Mv_N\rsb.
  \end{aligned}
\end{equation}
Substituting this expression back into the action we see that the factor \(C^2\) cancels with the
generalized background and we get
\begin{multline}
  S=-\frac{1}{4}\int d^5 x \sqrt{g}\lsb
  g^{MR}g^{NS}f_{MN}f_{RS}+4Gg^{zz}g^{NS}f_{zN}v_S+\right.\\+\left.
  2G^2g^{zz}g^{MN}v_Mv_N-2G^2g^{Mz}g^{Nz}v_Mv_N\rsb.
\end{multline}
Since we use the axial gauge for the field, i.e. \(v_z=0\), the second term in the new action
is simplified and the last term is eliminated,
\begin{equation}
  S=-\frac{1}{4}\int d^5 x \sqrt{g}\lsb
  g^{MR}g^{NS}f_{MN}f_{RS}+4Gg^{zz}g^{NS}v_S\pz v_N+2G^2g^{zz}g^{MN}v_Mv_N\rsb.
\end{equation}
The variation of the action reads
\begin{equation}
  \begin{aligned}
    \delta_v S=-\int d^5 x \sqrt{g}&\lsb
    g^{MR}g^{NS}f_{MN}\partial_R\delta v_S+Gg^{zz}g^{NS}v_S\pz \delta v_N+\right.\\&+\left.
    Gg^{zz}g^{NS}\delta v_S\pz v_N+G^2g^{zz}g^{NS}v_N\delta v_S\rsb.
  \end{aligned}
\end{equation}
Integrating by parts,
\begin{equation}
  \begin{aligned}
    \delta_v S=\frac{1}{\vk^2}\int d^5 x &\lsb
    -\partial_R\lb\sqrt{g}g^{MR}g^{NS}f_{MN}\rb-
    \pz\lb\sqrt{g}g^{zz}g^{NS}Gv_N\rb+\right.\\&+\left.
    \sqrt{g}g^{zz}g^{NS}G\pz v_N+
    \sqrt{g}g^{zz}g^{NS}G^2v_N\rsb\delta v_S,
  \end{aligned}
\end{equation}
the variation is further simplified to
\begin{equation}
  \begin{aligned}
    \delta_v S=\frac{1}{\vk^2}\int d^5 x &\lsb
    -\partial_R\lb\sqrt{g}g^{MR}g^{NS}f_{MN}\rb-
    \pz\lb\sqrt{g}g^{zz}g^{NS}G\rb v_N+\right.\\&+\left.
    \sqrt{g}g^{zz}g^{NS}G^2v_N\rsb\delta v_S.
  \end{aligned}
\end{equation}
The equation of motion follows
\begin{equation}
  -\partial_R\lb\sqrt{g}g^{MR}g^{NS}f_{MN}\rb-
  \pz\lb\sqrt{g}g^{zz}g^{NS}G\rb v_N+
  \sqrt{g}g^{zz}g^{NS}G^2v_N=0.
\end{equation}
The expansion of the first term results in
\begin{equation}
\label{v_antisym_int_exp}
  \begin{aligned}
    &\partial_R\lb\sqrt{g}g^{MR}g^{NS}f_{MN}\rb=
    \\&
    \partial_\rho\lb\sqrt{g}g^{\mu\rho}g^{\nu S}f_{\mu\nu}\rb+
    \partial_\rho\lb\sqrt{g}g^{\mu\rho}g^{zS}f_{\mu z}\rb+
    \pz\lb\sqrt{g}g^{zz}g^{\nu S}f_{z\nu}\rb.
  \end{aligned}
\end{equation}
Here the first contribution is
\begin{equation}
  \begin{aligned}
    \partial_\rho\lb\sqrt{g}g^{\mu\rho}g^{\nu S}f_{\mu\nu}\rb&=
    \frac{R}{z}\partial_\rho\lb\eta^{\mu\rho}\eta^{\nu\sigma}\partial_\mu v_\nu\rb-
    \frac{R}{z}\partial_\rho\lb\eta^{\mu\rho}\eta^{\nu\sigma}\partial_\nu v_\mu\rb=\\&=
    \frac{R}{z}\partial_\mu\partial^\mu \eta^{\nu\sigma}v_\nu-
    \frac{R}{z}\eta^{\nu\sigma}\partial_\nu\partial_\rho v^\rho=
    \frac{R}{z}\partial_\mu\partial^\mu \eta^{\nu\sigma}v_\nu.
  \end{aligned}
\end{equation}
In the last step, we used the usual Lorentz condition $\partial_\rho v^\rho=0$. With this condition, the term~\eqref{v_antisym_int_exp}
is simplified to
\begin{equation}
      \partial_R\lb\sqrt{g}g^{MR}g^{NS}f_{MN}\rb = \frac{R}{z}\partial_\mu\partial^\mu \eta^{\nu\sigma}v_\nu-
      \pz\lb\frac{R}{z}\pz\eta^{\nu\sigma}v_\nu\rb.
\end{equation}
Substituting the last two expressions into the equation of motion we get
\begin{equation}
  -\frac{R}{z}\partial_\mu\partial^\mu \eta^{\nu\sigma}v_\nu+
  \pz\lb\frac{R}{z}\pz\eta^{\nu\sigma}v_\nu\rb-
  \pz\lb\sqrt{g}g^{zz}g^{NS}G\rb v_N+
  \sqrt{g}g^{zz}g^{NS}G^2v_N=0,
\end{equation}
or, multiplying by \(z/R\) and using again $v_z=0$,
\begin{equation}
  \lsb-\partial_\mu\partial^\mu+
  z\pz\lb\frac{\pz}{z}\rb+z\pz\lb\frac{G}{z}\rb-G^2\rsb\eta^{\nu\sigma}v_\nu=0.
\end{equation}
We thus arrived at the equation~\eqref{v_simple_simplified_eom}. This proves that the considered generalized
background is fully equivalent to the infrared $\mathcal{O}(z^2)$ correction
to the 5D mass.

The analysis for symmetric vector case can be generalized to the case of tensor fields. As we discussed in the
Section 4.2, for spin \(J>1\) only the SW\(^+\) model can be generalized, and it has
the following action,
\begin{equation}
  S=\frac{1}{2}\int d^5x\sqrt{g}e^{|c|z^2}M^2(-b,J-1,-|c|z^2)\partial^M\Phi^{M_1\dots M_J}
  \partial_M\Phi_{M_1\dots M_J}.
\end{equation}
We follow the same procedure as above. First, we introduce a field redefinition to
eliminate the generalized background,
\begin{equation}
  \Phi_{M_1\dots M_J}\equiv C \phi_{M_1\dots M_J}, \quad C\equiv e^{-|c|z^2/2}M^{-1}(-b,J-1,-|c|z^2).
\end{equation}
After introducing a familiar notation,
\begin{equation}
  G\equiv \lb -1+2\frac{M'}{M}\rb |c|z,
\end{equation}
we can write the field \(z\)-derivative as
\begin{equation}
  \pz\Phi_{M_1\dots M_J}=C\pz \phi_{M_1\dots M_J}+CG\phi_{M_1\dots M_J}.
\end{equation}
Thus, the action for \(\phi_{M_1\dots M_J}\) is
\begin{multline}
  S=\frac{1}{2}\int d^5x\sqrt{g}\lsb
  g^{MN}g^{M_1N_1}\dots g^{M_JN_J}\partial_M\phi_{M_1\dots M_J}\partial_N\phi_{N_1\dots N_J}+\right.\\\left.
  g^{zz}g^{M_1N_1}\dots g^{M_JN_J}\lb 2G\phi_{M_1\dots M_J}\pz\phi_{N_1\dots N_J}+
  G^2\phi_{M_1\dots M_J}\phi_{N_1\dots N_J}\rb\rsb.
\end{multline}
After performing the variation of the action in the same way as in the vector case,
substituting various metric-related factors we get
\begin{equation}
  \sqrt{g}g^{MN}g^{M_1N_1}\dots g^{M_JN_J}=\frac{R^{3-2J}}{z^{3-2J}}
  \eta^{MN}\eta^{M_1N_1}\dots\eta^{M_JN_J},
\end{equation}
\begin{equation}
  \sqrt{g}g^{zz}g^{M_1N_1}\dots g^{M_JN_J}=-\frac{R^{3-2J}}{z^{3-2J}}
  \eta^{M_1N_1}\dots\eta^{M_JN_J},
\end{equation}
and multiplying by \(z^{3-2J}/R^{3-2J}\) we obtain the equation of motion
\begin{equation}
  \lsb-\partial_\mu\partial^\mu+
  z^{3-2J}\pz\lb\frac{\pz}{z^{3-2J}}\rb+
  \pz G-(3-2J)\frac{G}{z}-G^2\rsb
  \phi^{N_1\dots N_J}=0.
\end{equation}
Kummer's function \(M(-b,J-1,-|c|z^2)\) is a solution to the differential equation
\begin{equation}
  |c|z^2\frac{M''}{M}=(J-1+|c|z^2)\frac{M'}{M}+b
\end{equation}
After the same transformations that we did before when we used this equation, we arrive
at the final version of the equation of motion,
\begin{equation}
  \lsb\partial_\mu\partial^\mu-
  z^{3-2J}\pz\lb\frac{\pz}{z^{3-2J}}\rb+4b|c|-|c|(2-2J)+c^2z^2\rsb
  \phi^{N_1\dots N_J}=0.
\end{equation}
This equation represents a particular case of Eq.~\eqref{38} with \(m_\text{eff}^2R^2=c^2z^4+|c|z^2(4b-2+2J)\).

Finally we notice that the action used in this procedure is a
special case of a more general action Eq.~\eqref{genJ}, it is straightforward to extend the analysis above to
the scalar \(J=0\) case replacing (as it is dictated by the results of Section~4.1) the Kummer's function $M$ by the Tricomi one $U$.

\section*{Appendix C}
\addcontentsline{toc}{abcd}{\bf Appendix C}

\renewcommand\theequation{C.\arabic{equation}}
\setcounter{equation}{0}

In this paper, we encounter many times the differential equation
\begin{equation}
\label{hg_eq}
  \psi''(y)=\lb \frac{a_1}{y^2}+b_1+c_1y^2\rb\psi(y),
\end{equation}
for various real parameters \(a_1\), \(b_1\) and \(c_1\).
This equation looks simple to solve with the help of popular software systems like Mathematica or Maple
but for arbitrary parameters the produced results are not always trustworthy. In order to escape a confusion
and better understand what is going on with the solutions at different parameters, below we find
the exponentially decreasing and increasing solutions analytically.

In the limit \(y\to\infty\), we have the following approximate equation
\begin{equation}
  \psi''\approx c_1y^2 \psi,
\end{equation}
which has the approximate solution
\begin{equation}
  \psi=Ce^{s\sqrt{c_1}y^2/2},\quad s=\pm1,
\end{equation}
where \(C\) is an arbitrary constant. Since we may be interested both in
solutions that are convergent at \(y\to\infty\) as well as divergent solutions, the sign
of the exponent can be either plus or minus. Based on this observation we use the
following ansatz,
\begin{equation}
  \psi(y)=e^{sc_2y^2/2}\sum_{j=0}^\infty t_j y^{j+l},
\end{equation}
where \(c_2\equiv\sqrt{c_1}\), and \(s\equiv\pm 1\) depending on the convergence
requirements at \(y\to\infty\). After substituting into the Eq.~\eqref{hg_eq} and simple
manipulations we get
\begin{equation}
  sc_2\sum_{j=0}^\infty\lb 2j+2l+1-\frac{b_1}{sc_2}\rb t_j y^{j+l}+
  \sum_{j=0}^\infty\lb(j+l)(j+l-1)-a_1\rb t_j y^{j+l-2}=0.
\end{equation}
Next, we change the index in the first sum \(j\to j-2\) to get the same power of \(y\) in
both sums
\begin{multline}
  sc_2\sum_{j=2}^\infty\lb 2(j-2)+2l+1-\frac{b_1}{sc_2}\rb t_{j-2} y^{j+l-2}+\\+
  \sum_{j=0}^\infty \lb(j+l)(j+l-1)-a_1\rb t_j y^{j+l-2}=0.
\end{multline}
Then we extract first two terms from the second sum to get the same starting index as in
the first sum
\begin{multline}
  sc_2\sum_{j=2}^\infty\lb 2(j-2)+2l+1-\frac{b_1}{sc_2}\rb t_{j-2} y^{j+l-2}+\\+
  \lb l(l-1)-a_1\rb t_0 y^{l-2}+
  \lb l(l+1)-a_1\rb t_1 y^{l-1}+\\+
  \sum_{j=2}^\infty \lb(j+l)(j+l-1)-a_1\rb t_j y^{j+l-2}=0.
\end{multline}
Since the powers of \(y\) are linearly independent, to satisfy the equation we need to make
each series coefficient equal to zero. The equality to zero of coefficients of two middle
terms results in the following two conditions,
\begin{equation}\label{conv_ind_eq}
  l(l-1)=a_1,
\end{equation}
\begin{equation}
  t_1=0.
\end{equation}
In general, we could have resolved the requirement the other way, by setting \(t_0=0\) and
\(l(l+1)=a_1\), but this is equivalent to redefining \(l\) to \(l+1\).

The equation \eqref{conv_ind_eq} is called indicial equation. Since $a_1>0$ (otherwise
the potential in the Eq.~\eqref{hg_eq} is not bounded from below), the two solutions of
Eq.~\eqref{conv_ind_eq} will have opposite sign. Depending on its roots we can get different
linearly independent solutions of the original differential equation~\eqref{hg_eq}. The most
complicated case is when the difference between the roots of the indicial equation is an
integer (see also the comments at the end of this section).

The remaining sums in our case provide us with recurrent relation between the series
coefficients
\begin{equation}
  t_j=-\frac{sc_2\lb 2(j-2)+2l+1-\frac{b_1}{sc_2}\rb}{(j+l)(j+l-1)-l(l-1)}t_{j-2}.
\end{equation}
We change the indices \(j\to k+2\),
\begin{equation}\label{recur}
  t_{k+2}=-\frac{sc_2\lb 2k+2l+1-\frac{b_1}{sc_2}\rb}{(k+2)(k+2l+1)}t_k.
\end{equation}
Next, we change the indices \(k\to 2n-2\) and continue applying the relation up until \(t_0\),
\begin{equation}
  t_{2n}=-\frac{sc_2\lb 4n+2l-3-\frac{b_1}{sc_2}\rb}{2n(2n+2l-1)}t_{2n-2}=
  \frac{(-1)^ns^nc_2^n\prod\limits_{k=1}^n\lb 4k+2l-3-\frac{b_1}{sc_2}\rb}{2^nn!\prod\limits_{k=1}^n(2k+2l-1)}t_0.
\end{equation}
Then using the rising factorials notation,
\begin{equation}
  x^{(n)}\equiv\prod\limits_{k=1}^n(x+k-1),
\end{equation}
we rewrite our expression as
\begin{multline}
  t_{2n}=\frac{(-sc_2)^n\prod\limits_{k=1}^n\lb k+\frac{l}{2}-\frac{3}{4}-\frac{b_1}{4sc_2}\rb}
  {n!\prod\limits_{k=1}^n\lb k+l-\frac{1}{2}\rb}t_0=\\=
  \frac{\lb\frac{l}{2}+\frac{1}{4}-\frac{b_1}{4sc_2}\rb^{(n)}(-sc_2)^n}
  {\lb l+\frac{1}{2}\rb^{(n)}n!}t_0.
\end{multline}
So the solution to the equation takes the form
\begin{equation}\label{hyper_sol_sum}
  \psi(y)=e^{sc_2y^2/2}y^l\sum_{n=0}^\infty
  \frac{\lb\frac{l}{2}+\frac{1}{4}-\frac{b_1}{4sc_2}\rb^{(n)}}
  {\lb l+\frac{1}{2}\rb^{(n)}n!} (-sc_2y^2)^n,
\end{equation}
where we also set \(t_0=1\), which is equivalent to dividing an undetermined integration
constant by \(t_0\). This power series is actually a generalized hypergeometric series,
which defines Kummer's function,
\begin{equation}
\label{M}
  M(a,b,z)={}_1F_1(a,b,z)\equiv\sum_{n=0}^\infty\frac{a^{(n)}}{b^{(n)}n!}z^n.
\end{equation}
Thus, the final form of the solution is (we also substitute \(c_2=\sqrt{c_1}\))
\begin{equation}
\label{solM}
\psi(y)=e^{s\sqrt{c_1}y^2/2}y^l
  M\lb\frac{l}{2}+\frac{1}{4}-\frac{b_1}{4s\sqrt{c_1}},l+\frac{1}{2},-s\sqrt{c_1}y^2\rb,
\end{equation}
where \(l\) is the solution of the indicial equation \eqref{conv_ind_eq}. If we are interested in
a solution converging at \(y\to0\) then we should pick the positive root of the
indicial equation; otherwise, we should choose the negative root.

If we want to have a solution that diverges at \(y\to\infty\) it is enough to set
\(s=+1\). However, to get a convergent at infinity solution, besides choosing \(s=-1\)
we need to take extra measures, since in principle the sum in \eqref{hyper_sol_sum} can be
infinite. The recurrent relation on series coefficients \eqref{recur} has the following
asymptotics at large \(k\),
\begin{equation}
  t_{k+2}\sim\frac{t_k}{k}.
\end{equation}
Thus, if the series is infinite we will have a divergent solution. A series can be finite
only if the series coefficients are equal to zero starting from some \(k_\text{max}\).
This can only happen if the numerator of \eqref{recur} is equal to zero. This gives us
an equation for \(k_\text{max}\) and other parameters,
\begin{equation}
  2k_\text{max}+2l+1-\frac{b_1}{sc_2}=0.
\end{equation}
After introducing a new parameter \(k_\text{max}=2n\) and recalling that \(c_2=\sqrt{c_1}\),
we get a condition
\begin{equation}\label{termination_cond}
  b_1=s\sqrt{c_1}(4n+2l+1),
\end{equation}
which has to be true in order to have a solution that is convergent at infinity.

For the sake of completeness, we should mention that the Kummer function in~\eqref{M} is
a solution of Kummer's equation,
\begin{equation}
\label{kummer}
  x\frac{d^2\omega}{dx^2}+(b-x)\frac{d\omega}{dx}-a\omega=0.
\end{equation}
The two independent solutions of second order differential equation~\eqref{kummer} can be
written in the form~\cite{res}
\begin{equation}
\label{solK}
\omega_1=M(a,b,x), \qquad \omega_2=x^{1-b}M(a+1-b,2-b,x).
\end{equation}
Unfortunately, these solutions are not defined at any real values of parameters:
$\omega_1$ is a solution so long as b is not an integer less than 1 while
$\omega_2$ is a solution so long as b is not an integer greater than 1~\cite{res}.
Since in our case $b=J-1$ with spin $J=0,1,2,\dots$, these exceptions become problematic.
It is known that one can build a solution defined at any integer $b$ by taking
the following linear combinations of $\omega_1$ and $\omega_2$,
\begin{equation}
\label{Tricomi}
U(a,b,x)=\frac{\Gamma(1-b)}{\Gamma(a+1-b)}\,\omega_1+\frac{\Gamma(b-1)}{\Gamma(a)}\,\omega_2.
\end{equation}
The function $U$ is called Tricomi function. This function formally is not defined at
integer $b$ but can be analytically continued to any integer $b$~\cite{res}.
The Tricomi function~\eqref{Tricomi} can be also uniquely determined
as the solution of~\eqref{kummer} satisfying the property~\cite{res},
\begin{equation}
\label{Tricomi2}
U(a,b,x)\sim x^{-a},\qquad x\rightarrow\infty.
\end{equation}
In the most of cases, the solutions $\omega_1$ and $U$ are linearly independent.
When they are not, it is known~\cite{res} that instead of $\omega_1$
one can use $\omega_2$ or
\begin{equation}
\tilde{\omega}_2 = M(a,b,x)\ln x+x^{1-b}\sum_{k=0}^\infty t_k x^k,
\end{equation}
as a second solution. This is a direct consequence of situations when the difference between roots
of the indicial equation is an integer.

Thus, the second solution for $\psi(y)$ can be written in terms of the Tricomi function with the same arguments as in~\eqref{solM},
\begin{equation}
\label{tricomi_sol}
\psi(y)=e^{-\sqrt{c_1}y^2/2}y^l
  U\lb\frac{l}{2}+\frac{1}{4}-\frac{b_1}{4s\sqrt{c_1}},l+\frac{1}{2},\sqrt{c_1}y^2\rb.
\end{equation}

\section*{Appendix D}
\addcontentsline{toc}{abcd}{\bf Appendix D}

\renewcommand\theequation{D.\arabic{equation}}
\setcounter{equation}{0}

The holographic models considered in this work are supposed to describe, in the first instance,
the Regge phenomenology of light mesons in the large-$N_c$ limit and in the chiral limit. For the sake of completeness of our exposition,
we will give a short survey on the linear Regge trajectories in the sector of light predominantly non-strange mesons (for which the chiral limit is the most applicable)
in which the quark-antiquark component of their Fock state seems to dominate according to the known data.
We should make a caveat from the very beginning that this issue is rather controversial in the literature. But we hope that even if some states
are identified incorrectly or missed, this would not change significantly the general picture presented below, thanks to the statistical
reasons.

\begin{figure}[!ht]
 \center{\includegraphics[width=0.9\linewidth]{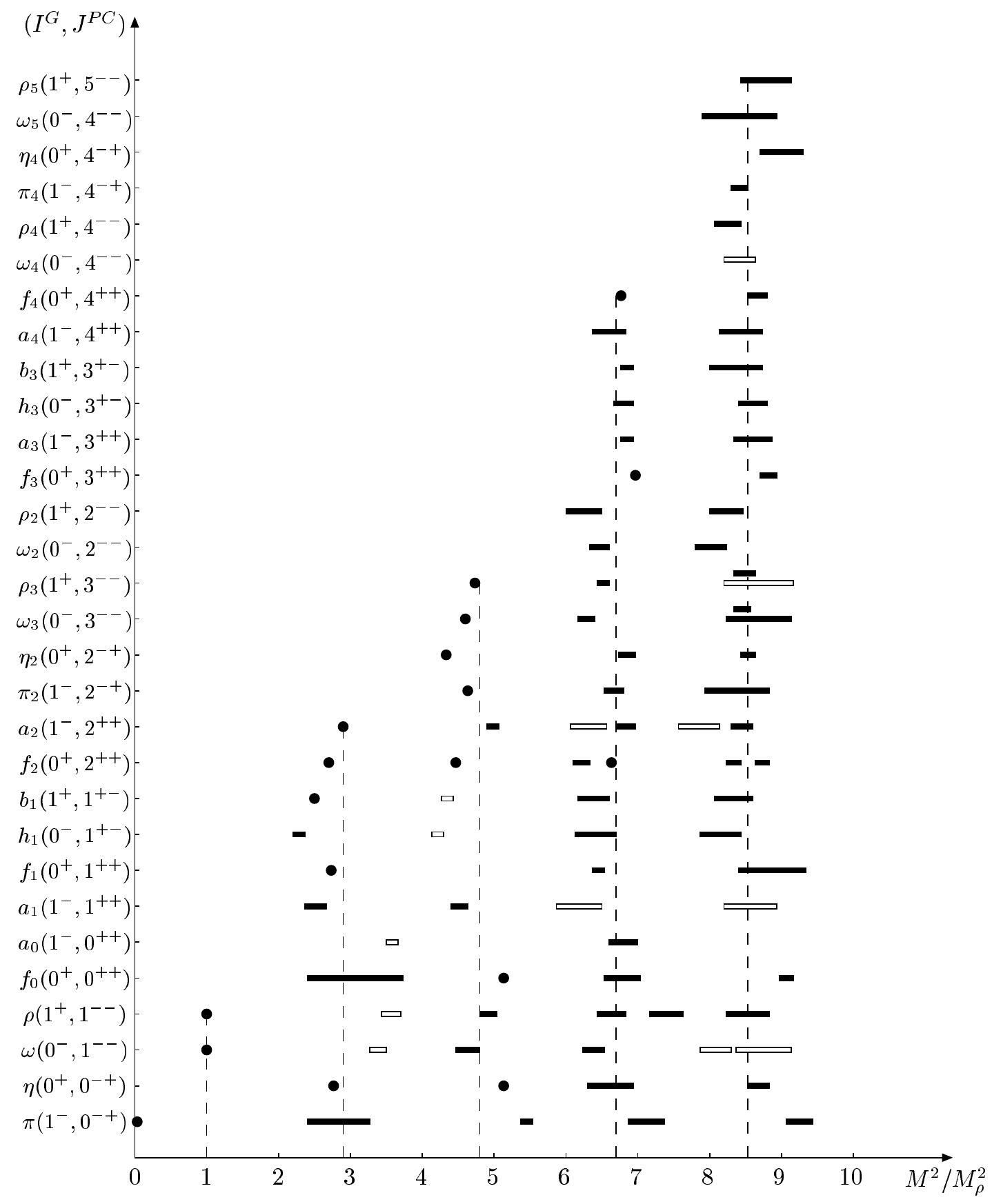}}
 \caption
{
\small
The spectrum of light non-strange quark-antiquark resonances in units of $M_{\rho(770)}^2$.
The data is taken from the Particle Data~\cite{pdg} and the Crystal Barrel
Collaboration~\cite{bugg} (many states in the last two towers).
The experimental errors are indicated.
The circles stay when the experimental errors are negligible on the plot. The dashed lines denote the positions of
mean (mass)$^2$ in each cluster of states. The open strips and circles are used
when the dominance of a quark-antiquark component is partly in question~\cite{pdg}.
Two states on the top of each tower do not have the parity doublets and form
the leading Regge trajectories (see the next plot).
}
 \label{BuggMesons}
\end{figure}

The spectrum of the mesons under consideration is displayed in Fig.~\ref{BuggMesons},
the relevant details are discussed in Refs.~\cite{phen3,phen4,phen5}.

\begin{figure}[!ht]
  \begin{minipage}[ht]{0.50\linewidth}
    \includegraphics[width=1\linewidth]{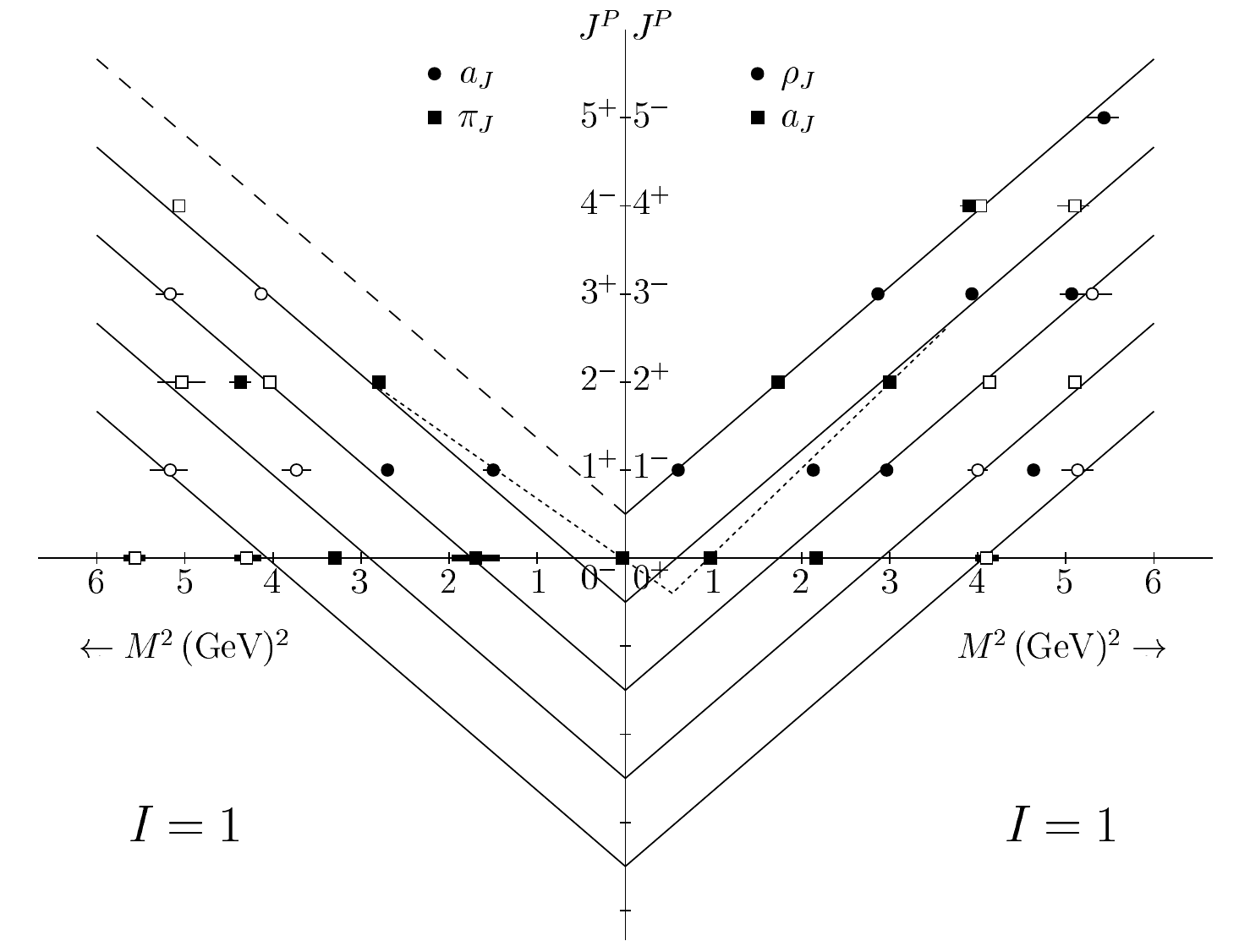} \\
  \end{minipage}
  \hfill
  \begin{minipage}[ht]{0.50\linewidth}
    \includegraphics[width=1\linewidth]{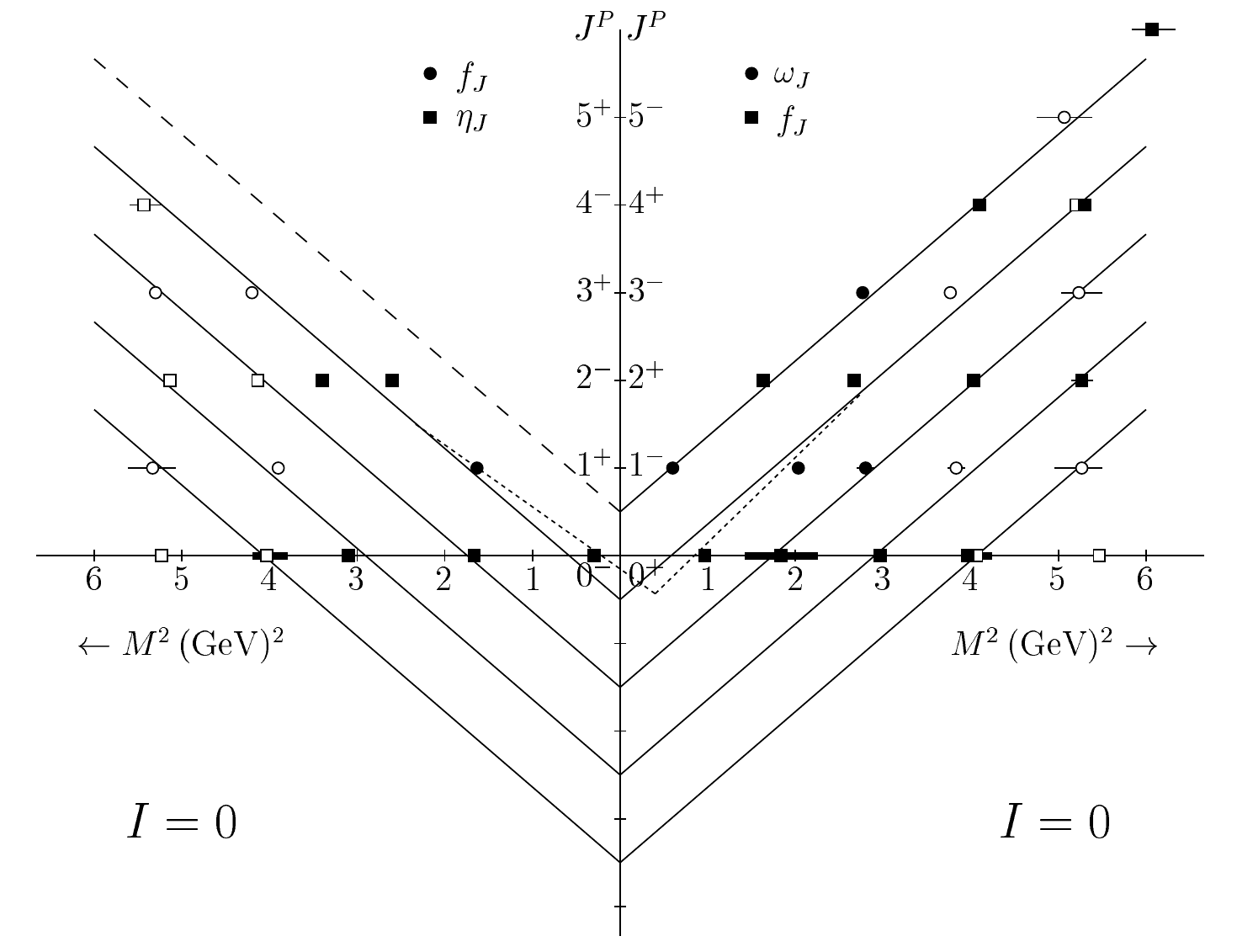} \\
  \end{minipage}
  \vspace{-0.8cm}
  \caption
{
\small
The Regge trajectories of isovector (left) and isoscalar (right) non-strange mesons.
Each trajectory actually consists of two degenerate Regge trajectories corresponding to even and odd spins.
The filled circles (squares) denote the states
contained in the Particle Data~\cite{pdg}. The open circles (squares) are the
resonances observed in the Crystal Barrel experiment~\cite{bugg}
(they are usually cited by the Particle Data in the section ``Further States'').
The parity doubling of trajectories is visualized via the reflection (the design of the plot is inspired by the MacDowell reflection symmetry
of fermion Regge trajectories leading to the parity doubling of the baryon trajectories, the relevant details are contained in Ref.~\cite{meson-baryon}).
The dashed line is a phantom image of the absent parity partner for the leading trajectory. The dotted line imitates the
distortion caused by the CSB.
}
  \label{macdowell}
\end{figure}

The global spectrum reveals two remarkable features: The clear-cut clustering of resonances near certain almost equidistant values of
energies squared and a specific
pattern of parity doubling --- the states lying on the leading Regge trajectories do not have parity partners
while all daughter Regge trajectories are parity doubled. The clustering and parity doubling was also observed in light baryons,
a review of the history of these observations and of proposed explanations is given in Ref.~\cite{parity}.
The parity doubling of meson Regge trajectories is geometrically visualized in Fig.~\ref{macdowell}.

It is well seen, in particular, that the leading $\rho$ and $\omega$ Regge trajectories have the intercept near $J=\frac12$.
For instance, if we use the masses of the most reliable states $\rho$ and $\rho_3$~\cite{pdg} on the $\rho$-trajectory to make the linear fit,
we obtain
\begin{equation}
\label{Jrho}
J_\rho\approx\frac{M^2}{1.13\,\text{GeV}^2}+0.52.
\end{equation}

\begin{figure}[!ht]
  \center{\includegraphics[width=0.9\linewidth]{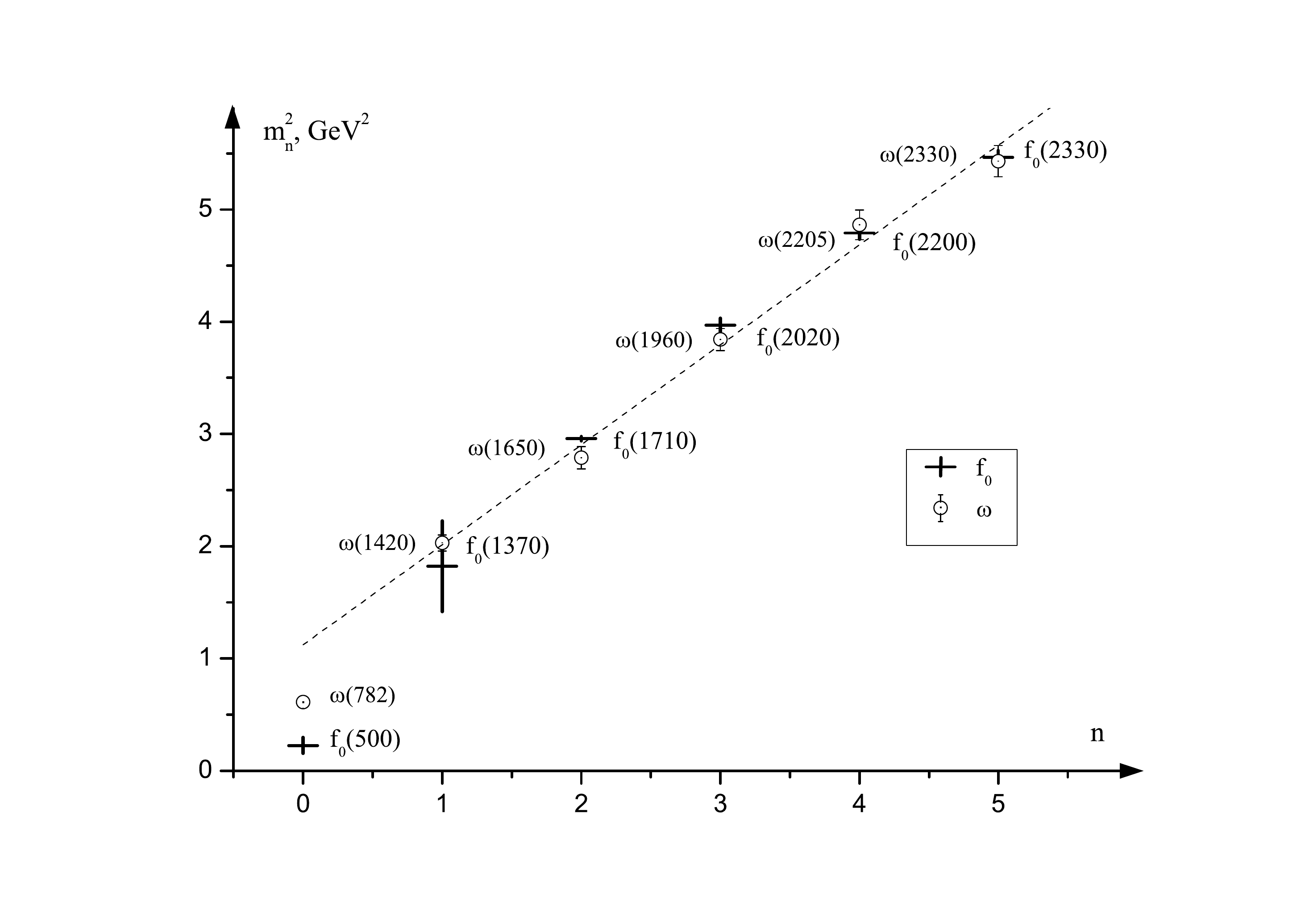}}
   \vspace{-0.9cm}
  \caption{\small A possible radial spectrum of light non-strange
$\omega$ (circles) and $f_0$ (crosses) mesons according to the Refs.~\cite{srAS,AK}.
The horizontal lines are drawn to distinguish better
the position of scalar resonances~\cite{pdg}.}
\label{OmegaF}
\end{figure}

The states possessing identical quantum numbers and lying on a (approximately equidistant) sequence of daughter Regge trajectories
form the ``radial'' trajectories, they represent direct analogues of the radial excitations in non-relativistic potential models.
The ground states below the CSB scale, about 1~GeV, lie appreciably below the positions predicted by the radial trajectories.
A couple of examples is shown in Fig.~\ref{OmegaF}

Taking the experimental masses from the Particle Data~\cite{pdg} one can make a global fit
of the data using the linear Regge ansatz and find the positions of clusters in Fig.~\ref{BuggMesons}.
 The first such analysis was performed in Ref.~\cite{phen3}, the result was (in GeV$^2$)
\begin{equation}
\label{cluster}
M^2_{\text{exp}}\approx1.14(N+0.54),
\end{equation}
where the integer $N=0,1,2$ enumerates the clusters.
One can further add the numerous resonances observed in the Crystal Barrel
experiment on the proton-antiproton annihilation in flight in the energy range 1.9--2.4~GeV, from which the spectrum of light non-strange mesons
was carefully extracted~\cite{ani,bugg}. The observed resonances populate the last two clusters in Fig.~\ref{BuggMesons}
 and allow to extend~\eqref{cluster} to $N=3,4$.
It turned out that the adding of many new resonances does not change noticeably the slope and intercept in~\eqref{cluster}~\cite{phen3}.
The averaged slope found in the compilation~\cite{bugg} coincides with the slope in~\eqref{cluster}.
It was proposed soon after (independently in Refs.~\cite{phen4,klempt,phenSV}) that the integer $N$ represents nothing but
an analogue of the principle quantum number in the Hydrogen atom,
\begin{equation}
\label{N}
N=L+n,
\end{equation}
where $n$ denotes the radial number and $L$ is the orbital momentum of quark-antiquark pair dictating also its spatial parity, $P=(-1)^{L+1}$.
All meson states in Fig.~\ref{BuggMesons}
can be classified with the help of $(L,n)$ assignment~\cite{phen4,phen5}. The resulting fit obtained in Ref.~\cite{phen4} was (in GeV$^2$)
\begin{equation}
\label{LplusN}
M^2\approx1.10(L+n+0.62).
\end{equation}
The slope from the fit~\eqref{LplusN} is used in the present work. Actually a very close slope for the orbital meson trajectory typically emerges
in other works, for instance, in the old phenomenological analysis of Ref.~\cite{Iachello:1991re}.

The relations like~\eqref{N} and~\eqref{LplusN} look non-relativistic since the orbital momentum $L$ cannot be separated from
the total angular momentum $J$ in a Lorentz-invariant way. It looks thus surprising why these relations work for light mesons which should
represent ultrarelativistic systems. It seems that this question is relative to the old question why the constituent quark model works.
Actually we still do not know how the strong coupling regime in QCD settles the physical degrees of freedom, they may well be non-relativistic ---
large effective quark and gluon masses are widely used in various phenomenological approaches.
In addition, we can always rewrite~\eqref{LplusN} in terms of the total momentum $J$, the price to pay will be the need to split~\eqref{LplusN}
into several relations for different sorts of mesons. For instance, the states lying on the leading Regge trajectory
have $L=J-1$ within the quark model. In this case, the spectrum~\eqref{LplusN} becomes close to the spectrum~\eqref{ls1b} in the next Appendix.

\section*{Appendix E}
\addcontentsline{toc}{abcd}{\bf Appendix E}

\renewcommand\theequation{E.\arabic{equation}}
\setcounter{equation}{0}

It is interesting to compare the experimentally motivated
spectrum~\eqref{LplusN} or~\eqref{cluster} with the old predictions of dual Veneziano like amplitudes derived in the framework
of the Regge theory~\cite{collins}, the relations~\eqref{ls1} and~\eqref{ls2} in the main text,
\begin{equation}
\label{ls1b}
m^2_{(\rho)}=a(n+J-1/2), \qquad J=1,2,\dots,
\end{equation}
\begin{equation}
\label{ls2b}
m^2_{(\pi)}=a(n+J), \qquad J=0,1,\dots.
\end{equation}
Such a comparison is relevant to the present work since the spectra like~\eqref{ls1b} and~\eqref{ls2b} emerge in the SW holographic models.

First of all, we should remind the reader the origin of spectra~\eqref{ls1b} and~\eqref{ls2b} from the Veneziano like amplitudes.
The original Veneziano dual amplitude for $\pi+\pi\rightarrow\pi+\pi$ scattering has the following general structure~\cite{collins},
\begin{equation}
\label{veneziano}
A(s,t)\sim B(1-\alpha(s),1-\alpha(t))=\frac{\Gamma(1-\alpha(s)) \Gamma(1-\alpha(t))}
{\Gamma(2-\alpha(s)-\alpha(t))}.
\end{equation}
The given ansatz possesses nice analytical properties dictated by the relativistic Regge theory, in particular,
the amplitude has poles at positive integer $\alpha=J$, where $J$ is interpreted as the spin of a resonance, according
to the Regge theory of complex angular momentum.

Consider resonances in the $s$-channel, where $s$ means the center-of-mass energy squared in a scattering process.
The requirement of linearity of Regge trajectories entails the linear relation,
\begin{equation}
\alpha(s)=\frac{s}{a}.
\end{equation}
The poles arise at
\begin{equation}
\frac{s}{a}=J,\qquad J=1,2,\dots.
\end{equation}
If the first pole at $J=1$ corresponds to $s=m_\rho^2$ we get the slope $a=m_\rho^2$.
The analytical properties of the amplitude~\eqref{veneziano} are preserved if we add the same terms but with shifted $\alpha$,
\begin{equation}
\alpha(s)\rightarrow \alpha(s)+n,\qquad n=0,1,2,\dots.
\end{equation}
The poles of resulting amplitude are then situated at
\begin{equation}
\label{venSP}
s=m^2(J,n)=m_\rho^2(J+n),
\end{equation}
giving rise to the leading, $n=0$, and daughter, $n=1,2,\dots$ Regge trajectories.
Exactly this spectrum was reproduced in the original SW holographic model~\cite{son2}. Such a spectrum is typical for various string approaches ---
this is not surprising as the Veneziano amplitude gave rise to the whole modern string theory.

The amplitude~\eqref{veneziano}, however, has a serious deficiency for the pion physics: $A(s,t)\neq 0$ as $s,t \to 0$.
Indeed, if pions are the Goldstone bosons, they must interact only via derivatives (an important consequence of the Goldstone theorem)
which become momenta in the momentum space, so at vanishing momentum the amplitude of $\pi\pi$ scattering must vanish as well.
The given soft pion theorem is known as the ``Adler self-consistency condition''~\cite{collins}.
This condition can be satisfied if we replace~\eqref{veneziano} by the following ansatz (the so-called ``Lovelace-Shapiro dual
amplitude''~\cite{collins}),
\begin{equation}
\label{lovelace}
A(s,t)\sim \frac{\Gamma(1-\alpha(s)) \Gamma(1-\alpha(t))}{\Gamma(1-\alpha(s)-\alpha(t))},
\end{equation}
supplemented by the condition
\begin{equation}
\label{adler}
\alpha(s)=\frac{1}{2}+ \frac{s}{a},
\end{equation}
and the same for $\alpha(t)$.
The obtained amplitude incorporates the spontaneous CSB but now does not correspond to any string theory ---
embedding the CSB into a string approach still remains an open problem.
Repeating the steps above ($s=m_\rho^2$ at $J=1$, etc.), we arrive at the spectrum
\begin{equation}
\label{lovSP}
s=m^2(J,n)=2m_\rho^2(J+n-1/2),
\end{equation}
which is the spectrum~\eqref{ls1b} for $a=2m_\rho^2$ or the spectrum~\eqref{ls1} in the main text.

As was shown in Ref.~\cite{avw}, the extension of this approach to more general reactions
$\pi+A\rightarrow B+C$ leads to the appearance of a quantization condition for Regge trajectories generalizing the
condition~\eqref{adler} and this results in emergence of the pion Regge trajectory with the spectrum~\eqref{ls2b},
in which $a=2m_\rho^2$ (the spectrum~\eqref{ls2} in the main text).

The discussion above refers to the exact chiral limit, $m_\pi^2=0$. With non-zero pion mass, we should
replace the Adler condition $\alpha(0)=\frac12$ in~\eqref{adler} by the condition
$\alpha(m_{\pi}^2)=\frac12$, i.e. we need to subtract the contribution of pion mass, $\alpha \to \alpha-\frac{m_{\pi}^2}{a}$,
in the r.h.s. of the condition~\eqref{adler}. This shifts the slope of all trajectories to a lower value~\cite{avw,avw2,collins},
\begin{equation}
\label{slope_a}
a=2(m_\rho^2-m_\pi^2).
\end{equation}
It should be noticed that the given shift improves the agreement with the mean experimental slope: Setting $m_\rho=0.769$~GeV
(the mass of neutral $\rho$-meson seen in the photoproduction of this resonance and in reactions with pions~\cite{pdg}) and $m_\pi=0.14$~GeV,
we get $a=2m_\rho^2\approx1.18$~GeV$^2$ in the chiral limit and $a=2(m_\rho^2-m_\pi^2)\approx1.14$~GeV$^2$ in the real world. The latter
value coincides with the fit~\eqref{cluster} and with the mean slope extracted in the compilation~\cite{bugg}.

Finally we should emphasize that the spectrum~\eqref{ls1b} and~\eqref{ls2b}, despite all its theoretical elegance, does not describe
the behavior of experimental spectrum of light mesons, rather it yields a reasonable description only for the states corresponding to $n=0$,
i.e. belonging to the leading Regge trajectories. This was demonstrated in the phenomenological analysis of Ref.~\cite{phen2}.
Actually the need to find a working relation lead to the model~\eqref{LplusN}.

\end{document}